\documentclass[%
 reprint,
superscriptaddress,
aps,pra,
showkeys,
 amsmath,amssymb,
longbibliography
]{revtex4-2}

\IfFileExists{newtxtext.sty}
 {\usepackage{newtxtext,newtxmath}}
 {\IfFileExists{stix.sty}
 {\usepackage{stix}}
 {\IfFileExists{mathptmx.sty}
 {\usepackage{mathptmx}}{}}}

\usepackage{graphicx}% Include figure files
\usepackage{array} % better tabular
\newcolumntype{P}[1]{>{\centering\arraybackslash}p{#1}}
\newcolumntype{M}[1]{>{\centering\arraybackslash}m{#1}}
\usepackage{dcolumn}% Align table columns on decimal point
\usepackage{bm}% bold math
\usepackage{accents}
\usepackage{multirow}
\usepackage{hyperref}% add hypertext capabilities
\usepackage{braket}

\usepackage[acronym]{glossaries}
\makenoidxglossaries

\newacronym{QKD}{QKD}{Quantum Key Distribution}
\newacronym{QBER}{QBER}{quantum bit error rate}
\newacronym{TFQKD}{TF-QKD}{Twin-Field Quantum Key Distribution}
\newacronym{PLL}{PLL}{phase-locked loop}
\newacronym{SNR}{SNR}{signal-to-noise ratio}
\newacronym{EuroQCI}{EuroQCI}{European Quantum Communication Infrastructure}
\newacronym{SPAD}{SPAD}{Single Photon Avalanche Diode}
\newacronym{SNSPD}{SNSPD}{Superconductive Nanowire Single Photon Detector}
\newacronym{SNS}{SNS}{Sending-or-Not-Sending}
\newacronym{SNS-AOPP}{SNS-AOPP}{Sending or not sending with actively odd-parity pairing}
\newacronym{CAL}{CAL}{Curty-Azuma-Lo}
\newacronym{ASE}{ASE}{Amplified Spontaneous Emission}
\newacronym{PLOB}{PLOB}{Pirandola-Laurenza-Ottaviani-Banchi}
\newacronym{BB84}{BB84}{Bennett-Brassard 1984}
\newacronym{SPD}{SPD}{Single Photon Detector}
\newacronym{MDIQKD}{MDI-QKD}{measurement-device-independent QKD}

\usepackage[usenames,dvipsnames]{xcolor}
\usepackage{siunitx}

\begin{document}

\title{Phase Noise in Real-World Twin-Field Quantum Key Distribution}

\author{Gianluca Bertaina}
\affiliation{Istituto Nazionale di Ricerca Metrologica, Strada delle Cacce 91, I-10135 Torino, Italy}
\author{Cecilia Clivati}
\affiliation{Istituto Nazionale di Ricerca Metrologica, Strada delle Cacce 91, I-10135 Torino, Italy}
\author{Simone Donadello}
\affiliation{Istituto Nazionale di Ricerca Metrologica, Strada delle Cacce 91, I-10135 Torino, Italy}
\author{Carlo Liorni}
\affiliation{Leonardo Labs, Quantum Technologies Lab, Via Tiburtina, km 12,400 – Rome – 00131 – Italy}
\author{Alice Meda}
\affiliation{Istituto Nazionale di Ricerca Metrologica, Strada delle Cacce 91, I-10135 Torino, Italy}
\author{Salvatore Virzì}
\affiliation{Istituto Nazionale di Ricerca Metrologica, Strada delle Cacce 91, I-10135 Torino, Italy}
\author{Marco Gramegna}
\affiliation{Istituto Nazionale di Ricerca Metrologica, Strada delle Cacce 91, I-10135 Torino, Italy}
\author{Marco Genovese}
\author{Filippo Levi}
\affiliation{Istituto Nazionale di Ricerca Metrologica, Strada delle Cacce 91, I-10135 Torino, Italy}
\author{Davide Calonico}
\affiliation{Istituto Nazionale di Ricerca Metrologica, Strada delle Cacce 91, I-10135 Torino, Italy}
\author{Massimiliano Dispenza}
\affiliation{Leonardo Labs, Quantum Technologies Lab, Via Tiburtina, km 12,400 – Rome – 00131 – Italy}
\author{Ivo Pietro Degiovanni}
\affiliation{Istituto Nazionale di Ricerca Metrologica, Strada delle Cacce 91, I-10135 Torino, Italy}

\begin{abstract}
The impact of noise sources in real-world implementations of Twin-Field Quantum Key Distribution (TF-QKD) protocols is investigated, focusing on phase noise from photon sources and connecting fibers. This work emphasizes the role of laser quality, network topology, fiber length, arm balance, and detector performance in determining key rates. Remarkably, it reveals that the leading TF-QKD protocols are similarly affected by phase noise despite different mechanisms. This study demonstrates duty cycle improvements of over a factor of two through narrow-linewidth lasers and phase-control techniques, highlighting the potential synergy with high-precision time/frequency distribution services. Ultrastable lasers, evolving toward integration and miniaturization, offer promise for agile TF-QKD implementations on existing networks. Properly addressing phase noise and practical constraints allows for consistent key rate predictions, protocol selection, and layout design, crucial for establishing secure long-haul links for the Quantum Communication Infrastructures under development in several countries.
\end{abstract}


\maketitle

\gls{QKD} protocols have the potential to revolutionize the cryptographic environment, with solutions that enable to share keys between distant parties, with security claims guaranteed by the laws of quantum mechanics \cite{Scarani_securitypracticalquantum_2009,BennettBrassard1984}, without assumptions on the computational power of the attacker. After almost 40 years of theoretical work, numerous proof of principle experiments and deployment of testbeds \cite{Lo2014, Pirandola2020, Peev2009, Martin2019, Chen2021}, nowadays the real objective is the integration of this technology in long-distance fiber networks already utilized for classical telecommunication~\cite{Sasaki_Fieldtestquantum_2011,Takesue_Experimentalquantumkey_2015,Avesani_Resourceeffectivequantumkey_2021,Zhang_deviceindependentquantumkey_2022,hajomer2023continuousvariable,ribezzo2023quantum}. It is well understood that the range of \gls{QKD} links is limited by the channel losses, with the link maximum key rate upper limited by the repeaterless secret-key capacity or \gls{PLOB} bound \cite{Pirandola_Fundamentallimitsrepeaterless_2017}, which is stricter than a previous upper bound result~\cite{Takeoka_SquashedEntanglementQuantum_2014}. Trusted nodes are used to extend the achievable range, a temporary solution waiting for true quantum repeaters \cite{Briegel1998} to become deployable in the field.

\gls{TFQKD} is a solution that has been proposed few years ago \cite{Lucamarini_Overcomingratedistance_2018} to mitigate the negative impact of channel loss and reach key rates beyond the \gls{PLOB} bound without the use of a trusted node. \gls{TFQKD} is a type of \gls{MDIQKD}\cite{Lo_MeasurementDeviceIndependentQuantumKey_2012} in which the parties Alice and Bob encode information in the properties of dim laser pulses that are sent through optical fibers to a central untrusted relay node, Charlie, where they undergo single-photon interference that overcomes security challenges related to real device imperfections~\cite{Meda_Quantifyingbackflashradiation_2017,Huang_Implementationvulnerabilitiesgeneral_2018}. 
An important assumption, that makes these protocols more complex to deploy than, e.g., time-bin encoded \gls{BB84}, is that optical pulses need to be phase-coherent when they are generated in distant locations and preserve coherence throughout the propagation to Charlie in spite of vibrations, seismic noise and temperature fluctuations encountered along the path. The first requirement was initially achieved by mutually phase-locking the photon sources in Alice and Bob, distributing reference-phase information through a service fiber link, while the second is addressed by interleaving the \gls{QKD} signals with bright reference pulses that probe the fiber to detect and compensate its noise and recover stable interference visibility as required for low-error operation \cite{Wang2019, Minder_Experimentalquantumkey_2019}. These approaches become less effective with long-distance links, reducing the actual duty cycle of the \gls{QKD} transmission. In recent years, other solutions have been proposed \cite{Pittaluga2021, Clivati_Coherentphasetransfer_2022}, 
based on dual wavelength transmission and active stabilization of the \gls{QKD} lasers and/or connecting fibers. 
Since the first proposals, many noise cancellation variants have been implemented, trading-off performance with equipment and infrastructural complexity, which is a concern in the quest for realizing deployed and operational \gls{QKD} networks at reasonable cost. Emerging laser integration technologies \cite{kelleher2023, loh2020, lee2013} and interferometric techniques for the fiber length stabilization can play a role in this challenge. In addition, strong synergy exists with the concurrent realization of network-integrated services for the distribution of accurate time and ultrastable optical frequencies at a continental scale \cite{clonets,clivati2023, schioppo2022,dierix2016}, also in the frame of European initiatives such as the \gls{EuroQCI} \cite{EuroQCI}. 
Recently, \gls{TFQKD} has been demonstrated up to very long distances of $800-\SI{1000}{\km}$~\cite{Wang_Twinfieldquantumkey_2022,Liu_ExperimentalTwinFieldQuantum_2023}. These impressive results have been obtained by using controlled low-loss fiber spools, high repetition rates, and ultra low-noise cryogenic superconducting single-photon detectors. In our work, we focus on setups which could be realistically deployed in the near future, partially employing existing infrastructure.

The present work aims at providing a general formalism to model relevant impairments occurring in real-world point-to-point \gls{TFQKD} implementations, as well as the impact of practical constraints such as the length imbalance between the arms of the interferometer, the quality of the employed stabilization laser, the adopted network topology and the characteristics of the channel. The results of this analysis are then used to evaluate the performance of different \gls{TFQKD} protocols in terms of key rate vs channel loss/length. Relevant information is condensed in a minimal set of parameters that enables tailoring the analysis to other practical cases and is useful in the design and performance optimization of \gls{TFQKD} in real-world scenarios, in view of the establishment of long-haul operational connections. While, for the sake of conciseness, we present simulation results varying a few of the relevant parameters, we provide an open source model \cite{bertaina_zenodo_2024} to ease fast comparison among different scenarios.

This article is organized as follows:
In Sec.~\ref{sec:introduction} we introduce the standard scheme of a \gls{TFQKD} setup. In Sec.~\ref{sec:protocols} we introduce two prominent \gls{TFQKD} protocols and discuss the role of phase noise. In Sec.~\ref{sec:noise} we describe the main sources of phase noise in a \gls{TFQKD} scheme and model their contribution in view of estimating the achievable key-exchange performances. In Sec.~\ref{sec:detectors} we discuss the relevant figures of merit in the detection part of the \gls{TFQKD} apparatus and their impact on the key rate. Having outlined the complete model, in Sec.~\ref{sec:simulations} we report the results of the simulation of \gls{TFQKD} key rates during standard operation (namely absence of attack) under different scenarios stemming from the choice of phase stabilization, the length imbalance, and the detector parameters. Finally, in Sec.~\ref{sec:conclusions} we draw our conclusions and consider future perspectives. For completeness, the Appendices contain more details on our model of the phase fluctuations spectrum for different scenarios, and a detailed recap of the simulated protocols and their parameters.

\section{Introduction to TF-QKD}\label{sec:introduction}

\begin{figure}[!tbp]
 \centering
 \includegraphics[width=0.95\columnwidth]{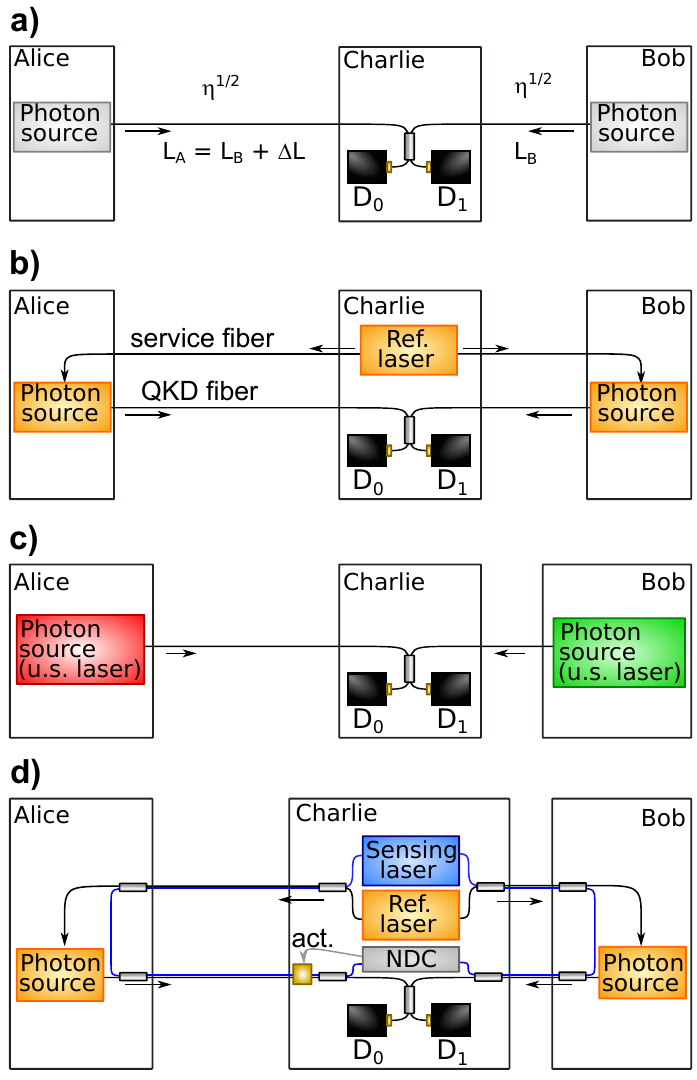}
 \caption{a) Principle scheme of a \gls{TFQKD} setup, characterized by total effective transmittance $\eta$, length of the interferometer arms $L_\text{A}$ and $L_\text{B}$, length imbalance $\Delta L$, and with D$_0$ and D$_1$ the single-photon detectors. b) Scheme of the \textit{Common-laser} approach to \gls{TFQKD}. c) Scheme of the \textit{Independent lasers} approach to \gls{TFQKD}, US laser: ultrastable lasers. d) Scheme of common-laser \gls{TFQKD} with fiber stabilization, NDC: noise detection and cancellation system, act.: actuator for fiber stabilization.}
 \label{fig:tfqkdsetup}
\end{figure}

A minimal model of a \gls{TFQKD} setup between two parties, Alice and Bob (abbreviated by A and B, respectively, and collectively indicated by \emph{they}) must take into account sources, channels, phase-coherence, detectors and protocol. See Figure~\ref{fig:tfqkdsetup}a.

\begin{itemize}
 \item \emph{Sources.} We consider attenuated laser sources, producing weak coherent states, characterized by intensities $\mu_i$ and phases $\varphi_i$, which can be both independently modulated in time by \emph{them} at a certain nominal clock rate $\nu_\text{s}$. 
 Phase noise of the sources is described in Sec.~\ref{sec:noise}. 
 \item \emph{Channels.} We consider optical fibers characterized by total attenuation (loss) $A_\text{T}$ or, equivalently, total transmittance $\eta=10^{-A_\text{T}/10}$, with $A_\text{T}$ measured in \si{\decibel}. The total loss $A_\text{T}=\alpha l + A_+$ is customarily given by a term proportional to the distance $l$, with (average) attenuation coefficient $\alpha$, plus additional losses $A_+$ due to instrumentation. Since in \gls{TFQKD} both of \emph{them} send signals to the auxiliary node Charlie (C), the relevant distances are $L_\text{A}$, on the segment AC between Alice and Charlie, and $L_\text{B}$, on the segment BC between Bob and Charlie. Without loss of generality, we assume $L_\text{A}\geq L_\text{B}$ and define the residual channel length imbalance as $\Delta L=L_\text{A}-L_\text{B}$. In order to maximize interference contrast in C, we balance the transmittances of the two segments in C, i.e. $\eta_\text{A}=\eta_\text{B}$, by assuming that a variable optical attenuator is introduced by Charlie on the CB segment. Therefore, the total effective transmittance between A and B is $\eta=(\eta_\text{A})^2$, corresponding to an effective total fiber length $2L_\text{A}$, which is in general larger than $L_\text{A}+L_\text{B}$. 

 \item \textit{Coherence.} Phase coherence between photon pairs generated at distant locations and interfered in Charlie is a peculiar prerequisite of \gls{TFQKD} and affects the overall \gls{QBER} and transmission duty cycle. 
 As a figure of merit for phase coherence, we introduce the variance of the phase fluctuations $\varphi$ observed at the detector in Charlie, $\sigma_\varphi^2$ 
 and quantify its contribution to the \gls{QBER} as~\cite{Clivati_Coherentphasetransfer_2022}
 \begin{equation}\label{eq:qberfromvar}
 e_\varphi=\int \sin\left(\frac{\varphi}{2}\right)^2 P(\varphi)\text{d}\varphi\approx \frac{\sigma_\varphi^2}{4} \;,
 \end{equation}
 where it is assumed that the phase fluctuations are Gaussian distributed.
 All protocols include dedicated hardware and routines to keep $\sigma_\varphi^2$ below a certain threshold during the key transmission, possibly introducing some dead-time and reducing the key rate. We model this effect by multiplying estimated key rates by a duty cycle $d=\tau_{\text{Q}}/(\tau_{\text{Q}}+\tau_{\text{PS}})$, namely the ratio between the maximum uninterrupted time $\tau_{\text{Q}}$ that is used in the quantum part of the key distribution protocol and the total time including the subsequent overhead spent for phase stabilization $\tau_{\text{PS}}$. $\sigma_\varphi^2$ and $\tau_{\text{Q}}$ are not independent: lower system phase noise allows for increasing $\tau_{\text{Q}}$. Conversely, $\tau_{\text{Q}}$ is upper-bounded by the time it takes for the system to reach a significant threshold for $\sigma_\varphi^2$. These aspects are treated in Sec.~\ref{sec:noise}.
 \item \emph{Detectors.} Charlie performs single-photon interference and detection on the two output branches of the interferometer, besides phase stabilization. The security proofs of the \gls{TFQKD} protocols guarantee that Charlie can be untrusted~\cite{Lucamarini_Overcomingratedistance_2018,Wang_Twinfieldquantumkey_2018,CurtyAzumaLo2019}, namely his actions can at worst deny \gls{QKD} operation, but not leak information to an attacker (Eve). Detectors are characterized by efficiency $\eta_{\text{D}}\leq 1$, that reduces the channels' transmission to $\hat{\eta}=\eta \eta_{\text{D}}$. Each detector is crucially characterized by dark counts per transmitted signal $p_{\text{DC}}$ given by the ratio of the dark count rate $P_{\text{DC}}$ and the clock rate $p_{\text{DC}}=P_{\text{DC}}/\nu_\text{s}$.
 The interferometer is affected by polarization misalignment $\theta$, which introduces an error $e_\theta=(\sin{\theta/2})^2$, whose impact depends on the used protocol. 
 Considerations about the characterization and improvement of detectors are made in detail in Sec.~\ref{sec:detectors}. 
 
 \item \emph{Protocols.} After the announcement of measurements by Charlie, Alice and Bob perform classical post-processing, exchanging information via a classical authenticated channel and estimating error rates from a small sample of the bits which are declared. To guarantee unconditional security, they perform error correction, to reconcile the raw bits, and privacy amplification, to remove the information possessed by Eve \cite{Shor_SimpleProofSecurity_2000,Gottesman_Securityquantumkey_2004}. 
 Error correction reduces the key size by an amount $f_{\text{EC}} Q H_2(E)$, where $Q$ is the total gain of the signals and $E$ is their total bit-flip \gls{QBER}. $H_2(p)=-p\log_2(p)-(1-p)\log_2(1-p)$ is the binary entropy and $f_{\text{EC}}$ is the inefficiency of error correction, that we customarily set to $f_{\text{EC}}=1.15$ \cite{Scarani_securitypracticalquantum_2009,Lucamarini_Overcomingratedistance_2018}.
 The actual amount of privacy amplification is specific to each \gls{QKD} implementation, as it depends on the detailed security analysis, protocol and parameters values. When the tagging argument \cite{Gottesman_Securityquantumkey_2004} and the decoy-state approach \cite{Lo_DecoyStateQuantum_2005,Ma_Practicaldecoystate_2005,Tamaki_Decoystatequantumkey_2016} are applied, typically the key length is reduced by an amount $\underaccent{\bar}{n}_1 H_2(\bar{e}_1^{\text{ph}})$, where $\underaccent{\bar}{n}_1$ is the estimated lower bound on the rate of single photon signal states at the detector and $\bar{e}_1^{\text{ph}}$ is the estimated upper bound on the single photon phase error rate. The total gain and single photon gains possibly include sifting factors depending on the specific protocol.
 Finally, the lower bound for the secure key per transmitted signal $\underaccent{\bar}{R}$ is to be multiplied by the source repetition rate $\nu_\text{s}$ and the duty cycle $d$ to obtain the total key rate.
 
 In the absence of quantum repeaters, the upper bound to secure key rate transmission per signal in a channel of total transmission $\eta$ is the \gls{PLOB} bound, namely the secret-key capacity of the channel $\text{SKC}_0=-\log_2(1-\eta)$, that scales as $1.44\eta$ at large losses~\cite{Pirandola_Fundamentallimitsrepeaterless_2017}. In \gls{TFQKD}, what matters is the branch with largest loss, corresponding to $\eta_\text{A}$. One has thus a much weaker dependence on the total distance, since $\eta_\text{A}=\eta^{1/2}$ when the AC and BC losses have been balanced. This enables overcoming the \gls{PLOB} bound and represents the most relevant achievement introduced by \gls{TFQKD}.
\end{itemize}

\section{TF-QKD protocols and role of laser phase noise}\label{sec:protocols}

In this Section we discuss two established \gls{TFQKD} protocols, \gls{SNS} and \gls{CAL}. 
In the original proposal of \gls{TFQKD} by Lucamarini et al.~\cite{Lucamarini_Overcomingratedistance_2018}, it was assumed that Eve cannot perform a Collective Beamsplitter attack, which relies on the knowledge of the global phase. This information is in fact leaked by the original protocol in order to match the phases chosen by Alice and Bob. Provably secure protocols in the assumption of coherent attacks were later introduced, like \gls{SNS} and \gls{CAL}, that rely on separating the communication in signal windows and decoy windows, used for precise parameter estimation. 
Both the \gls{SNS} (see \cite{Wang_Twinfieldquantumkey_2018, Yu2019,Xu_Sendingornotsendingtwinfieldquantum_2020} for details) and the \gls{CAL} (\cite{CurtyAzumaLo2019, Zhong2019, Grasselli2019} protocols employ weak coherent states in two complementary groups: phase mixtures and phase-definite states. Before reporting the key rates of the two protocols (and detailing their main steps in Appendix~\ref{app:SNSprotocol} and \ref{app:CALprotocol}), we generically comment on their reciprocity. Phase mixtures are phase-randomized coherent states of intensity $\mu$, which are seen by Charlie and Eve as statistical mixtures of number states $|n\rangle$ with Poissonian statistics $p_n^\mu=e^{-\mu}\mu^n/n!$. Since the intensity is weak, these mixtures mostly correspond to zero or one photon, and the security proofs relate those states to the eigenstates of the Z operator. These states are then said to belong to the Z basis and the corresponding measurement is related to photon counting. In contrast, by phase-definite states we mean coherent states that are to be employed in an interferometric measurement, which requires that a reference global phase is declared (either before or after Charlie's communication). These states are said to refer to the X basis, because they are related to the coherent superposition of zero and one photons in the security proofs.

The \gls{SNS} protocol uses the Z basis for encoding \cite{Wang_Twinfieldquantumkey_2018}, depending on the decision by Alice and Bob to send a number state or not. The decoy-state approach to phase error estimation is performed via interferometric measurement in the X basis. For applying standard decoy expressions, notice that the global phase is still randomized by Alice and Bob, but can be reconciled after Charlie's communication of measurement outcome. In contrast, in the \gls{CAL} protocol, the X basis is used for encoding \cite{CurtyAzumaLo2019}, and coherent states with two possible phases with $\pi$ difference are interfered in Charlie. Complementary, the counting Z basis is used in the decoy-state analysis, without need for reconciling the global phase. Both protocols remove the possible security issue in the original \gls{TFQKD} protocol, related to the need of revealing the global phase at each time window.

The secret key per transmitted signal for the original \gls{SNS} protocol is formulated in Eq.~\ref{SNSrate1} and we report here its formulation when \gls{SNS-AOPP} is used:
\begin{equation}
 \label{SNSrate2}
 \underaccent{\bar}{R} = p_\text{Z}^2 \left[{n'}_1 \left(1-H_2(\bar{e}_1^{' \text{ph}})\right)- f_{\text{EC}} n_\text{t} H_2(E'_\text{Z})\right]\,.
\end{equation}
While the description of the symbols in the equation is postponed to Appendix~\ref{app:SNSprotocol}, we discuss here the effect of phase fluctuations in the lasers and in the fiber on the error terms, which dominate the behavior of the key rate formula. $E'_\text{Z}$, the bit-flip error rate, is inherently independent of the phase fluctuations, since the encoding in the Z basis is phase-independent, given the previous considerations. The \gls{QKD} phase error rate $\bar{e}_1^{' \text{ph}}$, on the other hand, contains terms related to the phase fluctuations, since there the parties perform an interferometric measurement. 
In the \gls{CAL} protocol, the secret key per transmitted signal is estimated as Equation~\ref{CALrate1} of Appendix~\ref{app:CALprotocol}, that we report here:
\begin{align}
 &\underaccent{\bar}{R}_{X,k_\text{c} k_\text{d}}=\frac{1}{\nu_\text{s}}p_\text{XX}(k_\text{c},k_\text{d})\times \\ 
 &[ 1-f_{\text{EC}} H_2(e_{X,k_\text{c} k_\text{d}})-H_2(\text{min}\{
1/2,\bar{e}_{Z,k_\text{c} k_\text{d}} \}) ] \ . \nonumber
\end{align}
In this case, contrary to the \gls{SNS} protocol, the bit-flip error $e_{X,k_\text{c} k_\text{d}}$ is increased in the presence of phase fluctuations while $\bar{e}_{Z,k_\text{c} k_\text{d}}$ does not depend on them, since in the Z basis the states used for the decoy analysis are phase-randomized.

\section{Model for the laser phase noise}\label{sec:noise}

As discussed in Sec.~\ref{sec:protocols}, poor phase coherence between photon pairs interfering in Charlie increases the \gls{QBER} and reduces the final key rate, although its actual impact significantly depends on the used protocol.
In this Section, we explicitly derive decoherence effects for most common \gls{TFQKD} topologies and quantify the corresponding \gls{QBER}.

As a relevant metric to quantify decoherence, in Eq.~\eqref{eq:qberfromvar} we introduced the phase variance $\sigma_\varphi^2$ and its relation to $e_\varphi$, suggesting its relation to $\tau_\text{Q}$.
Indeed, the integration time that is relevant for calculating $\sigma_\varphi^2$ during the key transmission corresponds to the maximum uninterrupted transmission time $\tau_\text{Q}$.
To operationally quantify the relation between $\sigma_\varphi$ and $\tau_\text{Q}$, we will employ spectral analysis, as it gives more insight into relevant noise processes, simplifies calculations and is directly related to measurable quantities. We then introduce the noise power spectral density of a variable $y(t)$, $S_y(f) = \mathcal{F}[\mathcal{R}(y)]$, i.e. the Fourier transform of its autocorrelation function $\mathcal{R}(y)$, and will exploit its properties throughout the text \cite{papoulis}. According to the Wiener-Kintchine theorem, $\sigma_\varphi$ can be conveniently expressed in terms of the phase noise power spectral density $S_\varphi(f)$:
\begin{equation}
 \label{eq:jit_spec} \sigma_\varphi^2(\tau_\text{Q})= \langle\Delta^2 \varphi \rangle_{\tau_\text{Q}}= \int_{1/\tau_\text{Q}}^\infty S_\varphi(f) \,df \ . 
\end{equation}
$S_\varphi(f)$ is dominated by two contributions. Firstly, photons travel along telecommunication fibers, whose index of refraction $n$ and physical length $L$ change due to temperature, seismic and acoustic noise in the surrounding environment.
As a consequence, the phase accumulated by photons traveling through them changes over time.
A second contribution comes from the fact that the initial phases of twin photons generated in Alice and Bob cannot be perfectly matched, and the way this mismatch maps onto their interference in Charlie strictly depends on the experimental layout. We will now compare the most used topologies, providing relevant models for the various terms.

\subsection{Common-laser}

The typical way to ensure mutual phase coherence between Alice and Bob is to send them a common laser radiation, that can be used as a phase-reference to stabilize the local photon sources, so that they copy the phase of incoming light. This topology is depicted in Figure~\ref{fig:tfqkdsetup}b. The reference laser can be conveniently, though not necessarily, hosted by Charlie.
Incoming light in Alice and Bob is then a replica of the reference laser phase with additional noise due to propagation in the fiber (we assume that the stabilization of local laser sources to incoming light does not introduce noise).
The residual phase noise recorded upon interference in Charlie is (see Appendix~\ref{app:common-spectrum} for derivation):
\begin{equation}
\label{eq:noise_1laser}
S_\varphi(f) =
 4 \sin^2\left(\frac{2\pi f n \Delta L}{c}\right) S_\text{l,C}(f)
+4\left[S_\text{F,A}(f)+ S_\text{F,B}(f) \right] 
\end{equation}
where $S_\text{l,C}(f)$ is the noise of the reference laser, assumed to be at Charlie, and $S_\text{F,A}(f)$ ($S_\text{F,B}(f)$) is the noise of the fiber connecting Charlie to Alice (Bob).
The first term accounts for self-delayed interference of the reference laser: it vanishes if the propagation delays to Alice and Bob are equal, and progressively grows for larger length mismatches, with characteristic periodical minima at $f =k c/(2n\Delta L)$, $k$ integer, being $c$ the vacuum speed of light and $n=1.45$ the typical fiber refraction index.

Evidently, the quality of the reference laser impacts the residual noise of the interference. In this work, we consider representative cases of commercial, integrated diode lasers as well as state-of-the-art ultrastable lasers. 
Expressions and coefficients for the laser noise in these configurations are reported in Appendix~\ref{app:laser_phase_noise}, \ref{app:noise_coefficients} and Table~\ref{tab:coeffs}. Intermediate values are also possible, depending on the available technology and specific layout constraints.

The second term in Eq.~\ref{eq:noise_1laser} accounts for the fibers noise and depends on the environment where they are placed: metropolitan fibers affected by vehicle traffic and buildings vibrations show larger levels of noise than cables of similar length in country areas or seafloors. Similarly, suspended cables are found to be noisier than buried cables \cite{clivati2018}. Finally, the fiber noise may exhibit peaks around mechanical resonances of hosting infrastructures. 
Reasonable scaling rules hold for buried cables, which are the majority of those used for telecommunications on regional areas, under the assumption that the noise is uncorrelated with position and homogeneously distributed along the fiber. In this case, the noise can be assumed to scale linearly with the fiber length $L$ via an empirical coefficient $l$ \cite{williams2008}, and its expression includes a faster roll-off above a characteristic cut-off Fourier frequency $f_\text{c}'$:
\begin{equation}
\label{eq:freefibnoise}
 S_\text{F}(f, L) = \frac{l L}{f^2} \left(\frac{f_\text{c}'}{f+ f_\text{c}'}\right)^2 
\end{equation}
from which $S_\text{F,A}(f)=S_\text{F}(f, L=L_\text{A})$ and $S_\text{F,B}(f)=S_\text{F}(f, L=L_\text{B})$ follow. 
The multiplication factor 4 for these terms in Eq.~\ref{eq:noise_1laser} considers that noise is highly correlated for the forward and backward paths and adds up coherently. This is the actual scenario for parallel fibers laid in the same cable at Fourier frequencies $f \ll c/(nL_\text{A}),c/(nL_\text{B})$ . In other cases, this factor is reduced to 2 \cite{williams2008} and Eq.~\ref{eq:noise_1laser} provides thus a conservative estimation. Values for $l$ and $f_\text{c}'$ are derived in Appendix~\ref{app:noise_coefficients} and reported in Table \ref{tab:coeffs}.

Remarkably, the common laser approach can be conveniently implemented also in a Sagnac interferometer scheme \cite{Zhong2019,Zhong_Proofofprincipleexperimentaldemonstration_2021,Zhong_SimpleMultiuserTwinField_2022}, where two counter-propagating signals are launched from the central in opposite directions along the same fiber loop, reaching A and B, then looping-back to C. This configuration is inherently immune to length mismatches, hence the first term of Eq.~\ref{eq:noise_1laser} can be always neglected. Therefore the Sagnac-loop approach is particularly convenient in the case of multi-user ring networks. Nonetheless, our analysis is mainly focused on a different kind of architecture, specifically to point-to-point schemes, which can offer longer absolute reaches between two distant users when implemented with active phase noise cancellation techniques, as described subsequently.

\subsection{Independent lasers}

Another approach is based on independent lasers at the two terminals \cite{Wang_Twinfieldquantumkey_2018}, that are phase-aligned once at the start of the key transmission window and then let evolve freely for a finite amount of time, after which a new realignment is needed. This topology is sketched in Figure~\ref{fig:tfqkdsetup}c.
Following the same approach used to derive Eq.~\ref{eq:noise_1laser}, the phase noise of the interference signal in Charlie can then be modeled as:
\begin{equation}
\label{eq:noise_2lasers}
S_\varphi(f) =S_\text{l,A}(f)+S_\text{l,B}(f) 
+ S_\text{F,A}(f)+ S_\text{F,B}(f) 
\end{equation}
where the first and second terms describe the noise of the lasers at the two nodes, and the third and fourth terms indicate the noise of the fibers. In this topology, the fiber noise appears with coefficient 1, because the photon sources in A and B are independent and there is no round-trip of the radiation into connecting fibers (the auxiliary fiber has no role in this topology, besides classical communication services).
All relevant parameters are reported in Table \ref{tab:coeffs}. We note that again the quality of the used laser sources impacts the ultimate performances of the system, and overall higher instability and longer duty cycles could be achieved by employing lasers with superior phase-coherence.

When compared to the case of a common laser, discussed in the previous Section, the independent lasers approach is convenient from the point of view of the fiber network topology, since only a single fiber is required, and a round-trip is not necessary to distribute the reference laser. However, this introduces the drawback of the requirement of two ultrastable lasers, one in A and one in B. Indeed, here the laser quality plays an important role: in Eq.~\ref{eq:noise_2lasers} the laser noise terms sum up in Charlie, and they never cancel out as happens in the case of a common laser with balanced arm lengths as derived in Eq.~\ref{eq:noise_1laser}.
 
\subsection{Fiber noise cancellation strategies}

The most impacting term in Eqs.~\ref{eq:noise_1laser} and \ref{eq:noise_2lasers} is the fiber noise, which imposes the need for periodical phase-realignment, thus reducing the duty cycle $d$. 
This aspect can be addressed with a passive approach in the Sagnac-loop configuration, where to the first order the fiber noise at low frequencies is common-mode, self-compensating down to a limit that depends on the loop length \cite{Zhong2019,clivatiLargearea2013}. 
For long-range point-to-point connections, recent proposals \cite{Clivati_Coherentphasetransfer_2022,Pittaluga2021} suggested an alternative, usually referenced to as \textit{dual band stabilization}, that considerably relaxes the phase-realignment need, allowing to achieve $d>0.9$ exploiting high-bandwidth active phase noise cancellation techniques. 
This approach exploits an auxiliary sensing laser, traveling the same fiber as the single-photon packets, although at a detuned wavelength. Spectral separation techniques as those used in classical wavelength-division-multiplexing enable to detect interference signals produced by the sensing laser or the quantum signal on separate detectors with minimal cross-talks. The former is used to detect the fiber noise, while the latter performs the usual key extraction. 
First demonstrations of this approach were applied to the \textit{common-laser} setup (the corresponding scheme is depicted in Figure~\ref{fig:tfqkdsetup}d), and subsequently adapted to the independent-laser approach \cite{zhou2023}. Because there is no need to attenuate the sensing laser to the photon counting regime, its interference signal can be revealed by a classical photodiode with high \gls{SNR}. signal/noise ratio. The measurement and compensation of the differential phase noise between the two arms in C allow to overcome the limit given by the light travel delay, observed in alternative noise cancellation schemes \cite{williams2008,clivatiLargearea2013}. As a result, the fiber can be phase-stabilized in real-time and with a high bandwidth, e.g. by applying a suitable correction on an in-line phase or frequency modulator. 
Experimental demonstrations showed efficient rejection of the fibers noise, down to the limit:
\begin{equation}
\label{eq:stabfibnoise}
S_\text{F} (f, L) = \frac{(\lambda_\text{s}-\lambda_\text{q})^2}{\lambda_\text{s}^2} \frac{l L}{f^2}
\end{equation}
where $\lambda_\text{s}$ and $\lambda_\text{q}$ are the wavelengths of the sensing laser and quantum key transmission signal respectively, and the suppression factor $(\lambda_\text{s}-\lambda_\text{q})^2/\lambda_\text{s}^2$ is set by the fact that the fibers are stabilized based on the information from the former, while the quantum interference occurs at the latter \cite{Clivati_Coherentphasetransfer_2022}. 
Other reasons for deviation from the expected behavior may be short fiber paths that are not common between the two lasers (e.g. wavelength-selective couplers), whose fluctuations cannot be perfectly canceled. 
Advanced correction strategies can further suppress these contributions and ensure virtually endless phase stability \cite{Pittaluga2021}. Finally, detection noise of the sensing laser interference may represent the ultimate practical limit on very lossy links. This aspect is discussed in Appendix~\ref{app:noise_coefficients}. 

Interestingly, it can be seen that if the sensing laser is phase-coherent to the reference laser, i.e. $\varphi_\text{r}(t) \approx \varphi_\text{s}(t)$, the residual reference laser noise is canceled out together with the fiber noise. Phase-coherence between lasers separated by \SI{50}{\GHz} or \SI{100}{\GHz} (the minimum spectral separation that allows optical routing with telecom devices) can be achieved by locking multiple lasers to the same cavity or by phase-modulation-sideband locking. This concept has been further developed in \cite{zhou2023}, that successfully conjugates the independent-lasers approach with the fiber stabilization.

\section{Improving detection SNR}\label{sec:detectors}

The maximum communication distance is limited by the dark count rate $P_{\text{DC}}$ of the \glspl{SPD}, i.e. the intrinsic level of noise of the detector in the absence of any signal. 
$P_{\text{DC}}$ depends on the kind of single-photon detector used and on the operating conditions. At telecom wavelengths the most common solutions are InGaAs/InP \glspl{SPAD}, with either thermoelectric or Stirling cooling, and \glspl{SNSPD}. More details and less common technologies can be found in \cite{ceccarelli2021}.
\glspl{SNSPD} can reach dark count rates as low as $P_{\text{DC}} < $ \SI{0.01}{\Hz} \cite{pernice2012}, photon detection efficiency above \SI{90}{\percent} \cite{zadeh2017}, sub-\SI{3}{\ps} timing jitter \cite{Korzh2020} and dead-time below \SI{1}{\ns} \cite{Vetter2016}. These interesting properties come at the important cost of requiring cryostats capable to operate in the range \num{1}--\SI{4}{\K}, which is costly and adds technical limitations. Commercial solutions are now available that allow high-efficiency detection and quite low dark count level (typically $ \SI{10}{\Hz}$), but \glspl{SPAD} are still generally preferred for in-field applications, accepting lower general performance. Modern SPADs, working in gated mode, present photon detection efficiency around \SI{30}{\percent} \cite{scarcella2015}, timing jitter below \SI{70}{\ps} \cite{ceccarelli2021} and short dead-times, allowing to reach maximum count rates of more than \SI{500}{\MHz} with experimental devices (see, e.g., \cite{comandar2015}). Dark count rates vary considerably depending on the temperature of the sensor. Units that use thermoelectric cooling (around \SI{-40}{\degreeCelsius}) report values of hundreds of counts per second \cite{MPD-IR} or thousands. More effective Stirling coolers (reaching \SI{-100}{\degreeCelsius}) instead have $P_{\text{DC}} < $\SI{100}{\Hz} \cite{IDQ-IR}. Depending on the applications, other properties like maximum gating frequency, after-pulsing probability, back-flash probability and area of detection need to be taken into account. 
In a real world \gls{QKD} implementation, residual background photons due to the environment could be present in the dark fiber. 

It is important to reduce background photons at the same level or below the rate of the dark counts of the detectors. There are several sources of background photons. Photons may leak into the dark fiber from nearby fibers laid in the same cable, possibly hosting data traffic at wavelengths close to those used for the \gls{TFQKD} encoding. Moreover, in Sagnac-loop and time-multiplexed protocols, where strong classical signals at the quantum wavelength travel along the same fibers, single and double Rayleigh scattering can represent an important source of background photons \cite{Zhong2019,Chen_SendingorNotSendingIndependentLasers_2020}. The implementation of dual-band strategies for phase stabilization allows to strongly suppress these effects \cite{Pittaluga2021}, since phase stabilization relies on a different wavelength that can be suppressed by spectral filters. Nonetheless, advanced approaches to \gls{TFQKD} as those described before pose additional challenges. For instance, photons from the reference laser sent from Charlie to Alice and Bob through a separate fiber in the \textit{common-laser} scheme (\ref{sec:noise}) can be Rayleigh-scattered and evanescently couple to the fiber dedicated to the quantum transmission. Despite the probability of this process, combined to evanescent coupling, is small, the reference laser power must be carefully attenuated to ensure reliable referencing of slave lasers while keeping the background count rate suitably low. For example, the Rayleigh scattering effect becomes negligible when the power of the reference laser sent from Charlie to Alice and Bob through the respective service fibers is of the order of \SI{20}{\micro\watt} \cite{Clivati_Coherentphasetransfer_2022}. This guarantees sufficient reference signal for locking and regenerating the independent laser in Alice and Bob separated by hundreds of \si{\kilo\meter}, before being attenuated to the single photon level and encoded, giving negligible contribution to dark counts.
 
When dual-band noise detection and cancellation are considered, relevant sources of background photons in the quantum channel are the \gls{ASE} of employed laser sources and the Spontaneous Raman effect. \gls{ASE} noise from diode or fiber lasers considerably exceeds the spectral separation of the sensing and quantum lasers in a dual-band transmission. As the two co-propagate in the quantum fiber, efficient filtering would be required at the \glspl{SPD} in Charlie. Standard Bragg-grating filters employed in classical telecommunications have relevant drops in efficiency outside the range \SI{1300}{\nm}--\SI{1600}{\nm}, which may result in background \gls{ASE} photons to fall on the \glspl{SPD} and must therefore be complemented by dedicated equipment in \gls{TFQKD} setups. Raman scattering of the sensing laser propagating in the quantum fiber generates background photons on a broad spectrum, that extends to the \gls{QKD} wavelength channel. As such it cannot be efficiently filtered out, and the only mitigation strategy is again a careful adjustment of the launched sensing laser power to meet a condition where the Raman photon background remains negligible with respect to the signal. 
Instead of focusing on ultra-low-noise detectors, in our analysis we considered two realistic scenarios with different commercial detectors: one using best-in-class InGaAs/InP SPAD, with Stirling cooler and dark count rate from \SI{3}{\Hz} to \SI{60}{\Hz}, corresponding to a quantum efficiency respectively of \SI{10}{\percent} and \SI{25}{\percent};
the other adopting a {lower-noise and more efficient, but still commercially available} \glspl{SNSPD} with dark count rate of \SI{10}{\Hz} and quantum efficiency of \SI{90}{\percent}.

\section{Results for the simulation of realistic key rates}\label{sec:simulations}

\begin{figure*}[!tbp]
 \centering
 \begin{tabular}{cc}
 \includegraphics[width=1\textwidth]{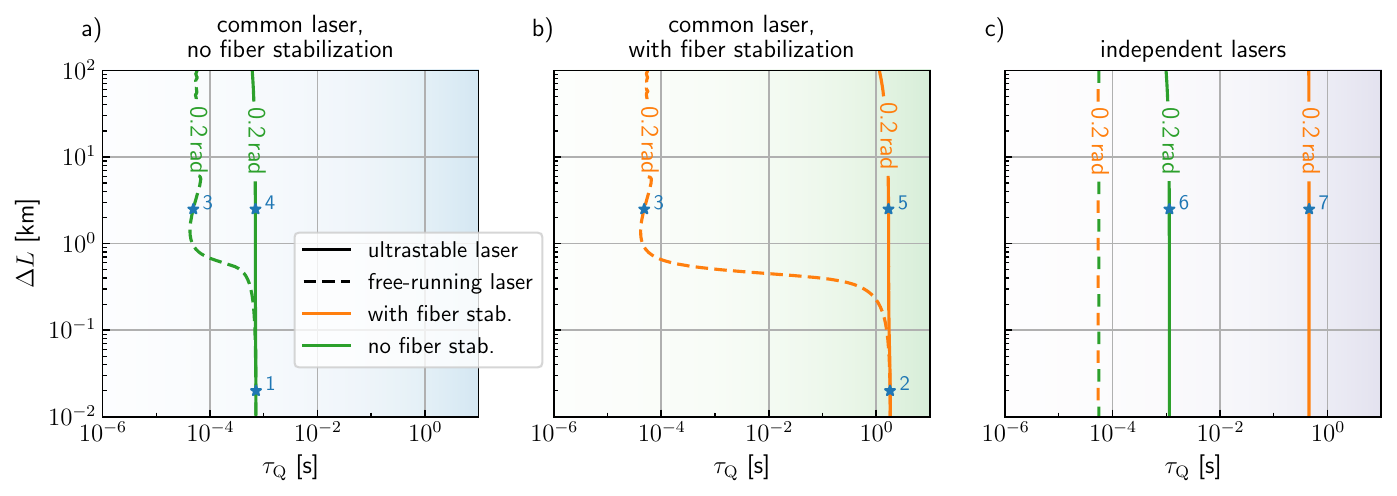}
 \end{tabular}
 \caption{Level curves at constant phase standard deviation $\sigma_\varphi$, calculated in the space of fiber length mismatch $\Delta L$ and integration time $\tau_\text{Q}$, for each possible combination of laser source configuration and fiber stabilization, at fixed shorter arm $L_\text{B}=\SI{100}{\km}$. The corresponding $\sigma_\varphi$ maps are reported in Figure~\ref{fig:noisescenarios}. The numbered points represent the specific scenarios of Table~\ref{tab:scenarios}, which are considered in the simulations.
 }
 \label{fig:sigma_unbalance}
\end{figure*}

Having discussed in the previous Sections the main experimental parameters that characterize the standard operation of the \gls{TFQKD} setup, in this Section we evaluate their impact on the expected key rates, focusing on the role of phase noise and of detector performance.
We consider various configurations that are possible for a \gls{TFQKD} layout, characterized by either common or independent sources, which are stabilized or not, with or without fiber stabilization, and with varying detector performance. For each combination, we calculate the phase noise and corresponding variance $\sigma_\varphi^2$ (Eq.~\ref{eq:jit_spec}). We fix an upper limit to tolerated phase fluctuations of $\sigma_\varphi=\SI{0.2}{\radian}$, that leads to $e_\varphi=0.01$ (Eq.~\ref{eq:qberfromvar}), which is a standard conservative value. The integration time $\tau_\text{Q}$ at which this threshold is achieved determines the duty cycle $d$, which is then used to evaluate the key rate. Notice that the duty cycle saturates to unity for integration times $\tau_\text{Q}\gg\tau_{\text{PS}}$ and that anyway $\tau_\text{Q}\gtrsim \SI{1}{\s}$ is probably unrealistic due to general realignment processes that are nevertheless to be performed. In particular, polarization drift given by fiber birefringence and thermal effects introduce polarization mismatch errors that must be corrected periodically. Experimental experience and literature show that these effects occurs on slow timescales, and polarization re-alignment procedures at low rates of a few \si{\hertz} are typically sufficient \cite{zhou2023,Li_TwinFieldQuantumKey_2023}. Due to frequency drifts of the free-running local oscillators used to clock encoders and decoders, typically time re-synchronization routines must be performed periodically every few seconds to maintain the required synchronization between the nodes.
Our approach of fixing the phase error threshold could be relaxed, in a more refined approach, by optimizing $\tau_\text{Q}$ and $\sigma_\varphi$ to maximize the key rate for each protocol and distances separately.

In Figure~\ref{fig:sigma_unbalance} we report the isolines matching $\sigma_\varphi=\SI{0.2}{\radian}$ as a function of $\tau_\text{Q}$ and fiber length mismatch $\Delta L=L_\text{A}-L_\text{B}$ at which such threshold is reached. As a reference, the length of the shorter interferometer arm $BC$ is considered constant and equal to $L_\text{B}=\SI{100}{\km}$. We observe that the most favorable configuration, with $\tau_\text{Q}$ exceeding \SI{1}{\s}, is reached with the configuration of a common cavity-stabilized laser, with fiber stabilization (panel b, solid line). This configuration is mostly insensitive to fiber length mismatch, over any reasonable range. On the contrary, if an unstabilized free-running laser is considered (dashed line), the mismatch causes $\tau_\text{Q}$ to drop rapidly below \SI{100}{\micro\second} for $\Delta L$ greater than a few hundred meters. This distance represents the crossover to a regime where the integrated laser noise exceeds the given phase noise threshold, and scales as the coherence length of the laser. The analogous configurations without fiber stabilization (panel a) show similar behaviors, although the highest $\tau_\text{Q}$ values are lower by more than $3$ orders of magnitude due to fiber noise. Finally, as expected, the configurations with independent laser sources (panel c) do not show any dependence on the fiber length mismatch. In particular, the configurations with free-running independent laser sources (dashed lines) correspond to very low $\tau_\text{Q}$ values. Conversely, when ultrastable lasers are considered, we observe a significant increase in $\tau_\text{Q}$, similar to the common laser case. Details of this analysis are described in Appendix~\ref{app:qbertau}.

\begin{table}[bt]
\centering
 \setlength{\extrarowheight}{2pt}
 \begin{tabular}{lccc}
  \hline
  Source configuration & Fiber stabilization & $\Delta L$ & Scenario \\
  \hline
  \multirow{3}{*}{Free-running common laser} & 
  \multicolumn{1}{c}{NO} & 
  \multicolumn{1}{c}{\SI{20}{\m}} &
  \multicolumn{1}{c}{1}\\
  \cline{2-4} &
  \multicolumn{1}{c}{YES} & 
  \multicolumn{1}{c}{\SI{20}{\m}} & 
  \multicolumn{1}{c}{2}\\
  \cline{2-4} &
  \multicolumn{1}{c}{ANY} & 
  \multicolumn{1}{c}{\SI{2.5}{\km}} &
  \multicolumn{1}{c}{3}\\
  \hline
  \multirow{2}{*}{Ultrastable common laser} & 
  \multicolumn{1}{c}{NO} & 
  \multicolumn{1}{c}{\SI{2.5}{\km}} & 
  \multicolumn{1}{c}{4}\\
  \cline{2-4} &
  \multicolumn{1}{c}{YES} & 
  \multicolumn{1}{c}{\SI{2.5}{\km}} & 
  \multicolumn{1}{c}{5}\\
  \hline
  \multirow{2}{*}{Ultrastable independent lasers} & 
  \multicolumn{1}{c}{NO} & 
  \multicolumn{1}{c}{ANY} & 
  \multicolumn{1}{c}{6}\\
  \cline{2-4} &
  \multicolumn{1}{c}{YES} & 
  \multicolumn{1}{c}{ANY} & 
  \multicolumn{1}{c}{7}\\
  \hline
 \end{tabular}
 \caption{Considered scenarios, marked as stars in Figure~\ref{fig:sigma_unbalance}, whose key rates are evaluated in Figures \ref{fig:keyrates_scenarios_SNSPD} and \ref{fig:keyrates_scenarios_SPAD}. They are characterized by source and fiber configurations and the length mismatch $\Delta L$.}
 \label{tab:scenarios}
\end{table}

To study the impact of these parameters on key rate, we now restrict our attention on the seven realistic scenarios listed in Table~\ref{tab:scenarios}, representing specific configurations and values of $\Delta L$, and we mark them as stars in Figure~\ref{fig:sigma_unbalance}. Since with just few hundred meters mismatch the impact of unstabilized laser source noise is significant~\cite{zhou2023}, in scenarios 3-7 we set a representative significant length mismatch $\Delta L=\SI{2.5}{\km}$, while for scenarios 1 and 2 we model a negligible mismatch of $\Delta L=\SI{20}{\m}$. Even in the most favorable configurations, $\tau_\text{Q}$ was limited to \SI{100}{\ms} to conservatively account for general realignment processes (polarization, time re-synchronization) required beyond this limit. 
Based on these scenarios, we simulate the key rates of the \gls{CAL} and \gls{SNS-AOPP} protocols, assuming a phase-synchronization overhead of $\tau_{\text{PS}}=1 \text{ms}$~\cite{Clivati_Coherentphasetransfer_2022}. As a reference, we consider a "realistic" \gls{PLOB} bound, where the transmission is effectively multiplied by the same detection efficiency $\eta_{\text{D}}$ employed in the simulations of the other protocols, and we also evaluate the key rate for the phase-based efficient \gls{BB84} protocol endowed with decoy states \cite{Hwang_QuantumKeyDistribution_2003,Wang_BeatingPhotonNumberSplittingAttack_2005,Lo_DecoyStateQuantum_2005,Ma_Practicaldecoystate_2005,Tamaki_Decoystatequantumkey_2016}. The latter employs the same decoy parameter estimation and channel and detector models as described in Appendix~\ref{app:SNSprotocol} and summarized in Appendix~\ref{app:decoy}, however assuming no role for phase noise. The parameters used in the simulations are described in Appendix~\ref{app:simparameters}. To analyze the impact of detector performance, we reproduce the scenarios considering either \glspl{SNSPD} or \glspl{SPAD}, as described in Sec.~\ref{sec:detectors}. 

\begin{figure*}[tp]
 \centering
 \includegraphics[width=1\textwidth]{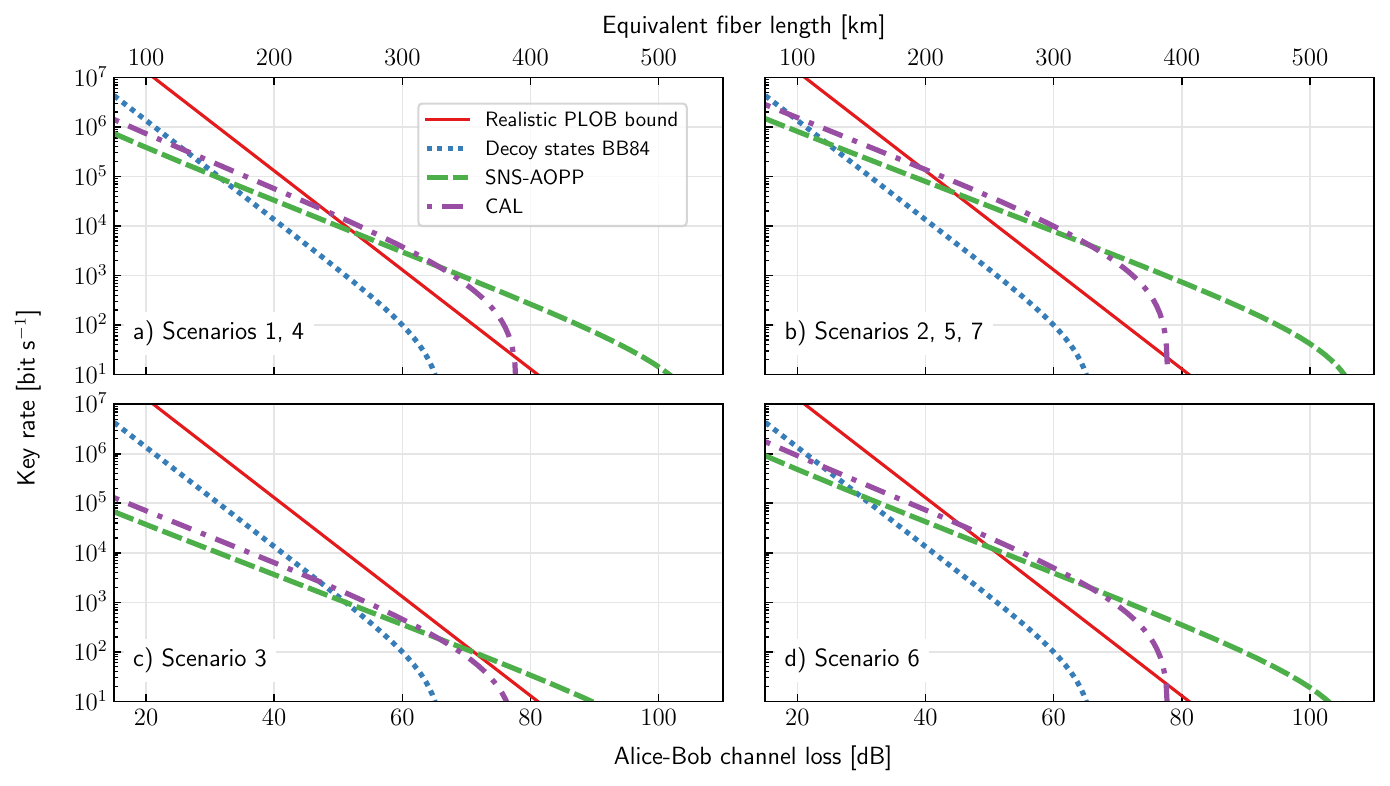}
 \caption{Simulated key rates of the \gls{BB84}, \gls{SNS-AOPP} and \gls{CAL} protocols in the scenarios described in Table~\ref{tab:scenarios}, with varying total loss and considering \glspl{SNSPD}. A reference \gls{PLOB} bound with effective attenuation is plotted. Panel a (b) reports simulations of scenario 1 (2), which are graphically indistinguishable from those of scenario 4 (5, 7).}
 \label{fig:keyrates_scenarios_SNSPD}
\end{figure*}

\begin{figure*}[bp]
 \centering
 \includegraphics[width=1\textwidth]{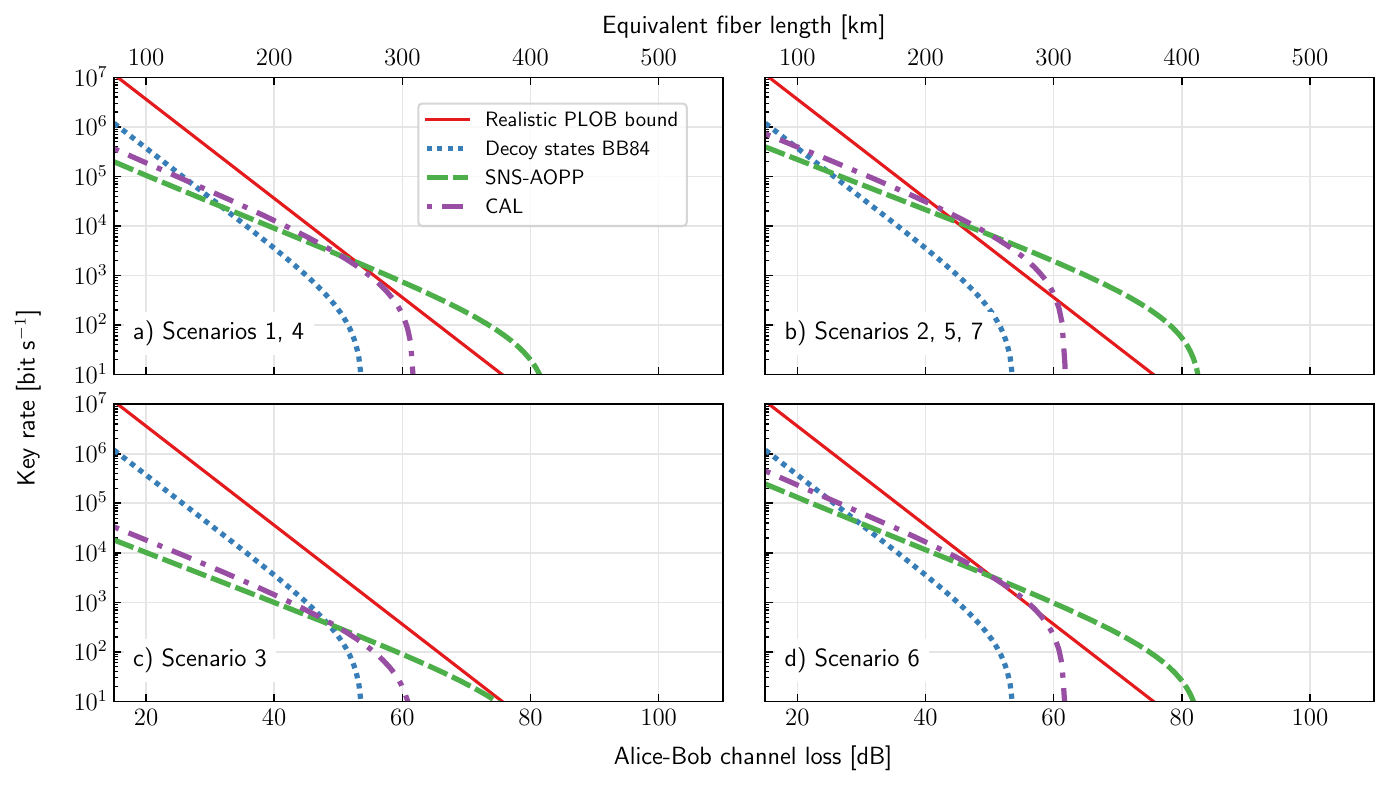}
 \caption{Key rates as in Figure~\ref{fig:keyrates_scenarios_SNSPD}, but considering representative \glspl{SPAD}.}
 \label{fig:keyrates_scenarios_SPAD}
\end{figure*}

Because of similar values of $\tau_\text{Q}$, some of the seven scenarios of Table~\ref{tab:scenarios} result in the same or very similar key rates, therefore we group them in four representative panels of Figure~\ref{fig:keyrates_scenarios_SNSPD}, where best performing representative \glspl{SNSPD} are considered.
These results show that unstabilized fibers are usually the largest contribution to decoherence and limit $\tau_\text{Q}$ to less than \SI{1}{\ms} (Scenarios 1 and 4, panel a, and scenario 6, panel d), except for the case when unstable lasers are used and no care is taken to match the optical paths' length. In this case, the residual self-delayed laser noise limits $\tau_\text{Q}$ to about \SI{50}{\micro\s} (Scenario 3, panel c). Using fiber and laser stabilization, either with a single common (Scenario 5, panel b) or a pair of lasers (Scenario 7, panel b), ensures $\tau_\text{Q} > $ \SI{100}{\ms} even in the presence of a large unbalance in the interferometer arms, with corresponding duty cycles approaching \SI{100}{\percent} and no impact on the \gls{QBER}. The only unstabilized laser scenario matching 5 and 7 in panel b is scenario 2, only because there is negligible length mismatch and the dominant noise is taken care of by fiber stabilization. In all scenarios, a prepare-and-measure approach features worse key rates than \gls{TFQKD}, for losses significantly larger than those typical of metropolitan networks.
In Figure~\ref{fig:keyrates_scenarios_SPAD}, the same scenarios are considered, but the detectors are best-in-class \glspl{SPAD}, as representative of more standard in-field setups. Quite generically, the increased dark count rate reduces the maximum reachable distance, while the lower efficiency reduces the key rates for any distance. Besides these very noticeable differences with respect to Figure~\ref{fig:keyrates_scenarios_SNSPD}, the considerations that we made pertaining to the role of phase noise are unaffected.

In the simulations shown here and in Appendix~\ref{app:SNSprotocol}, \ref{app:CALprotocol}, we recall that we make the assumption that asymmetric channels are treated by adding losses $A_+$ in the channel with higher transmittance. Optimized protocols have been proposed for both \gls{SNS} 
\cite{Hu_Sendingornotsendingtwinfieldprotocol_2019}
and \gls{CAL} \cite{Grasselli_Asymmetrictwinfieldquantum_2019} \cite{Wang2019,Wang_Simplemethodasymmetric_2020} for such asymmetric case. The intensities of the signals and decoy intensities at the two transmitters, among the other parameters, are independently optimized, achieving a higher key rate than with adding losses. We point out that these approaches only solve the problem of mismatched intensities at Charlie. Fiber length mismatches will still introduce problems related to distributed laser coherence, as discussed in detail in Sec.~\ref{sec:noise}, requiring the addition of fiber spools. Usually the need of fiber spools can be avoided in the case of Sagnac fiber-loop networks \cite{Wang_Simplemethodasymmetric_2020,Zhong_Proofofprincipleexperimentaldemonstration_2021,Zhong_SimpleMultiuserTwinField_2022}, where the length unbalances are intrinsically compensated. Conversely, the mismatch effects are evident when using free-running lasers in point-to-point schemes, where the fiber lengths should be matched at the order of \SI{100}{\meter} to obtain an optimal transmission window. Although being feasible in laboratory, this can represent an important limit to implement \gls{TFQKD} in realistic networks, where the fibers and the communication nodes are deployed according to different criteria, typically interconnecting inhabited towns along existing infrastructures and depending on the territory topology. This way, the unbalance between nodes is typically of several \si{\kilo\meter}, which in principle can be compensated with the addition of dedicated fiber spools, possibly in the service channels. Balancing the interferometer arm lengths by using fiber spools needs preliminary calibrations, requires additional hardware to be installed at the telecom shelters, and limits the flexibility in terms of possible dynamic network reconfiguration. All these factors pose constraints to the fiber providers, which are not negligible in production environments. Our results show that such requirements can be strongly relaxed by taking advantage of the reduced phase noise resulting from ultrastable lasers and fiber noise cancellation techniques, improving the applicability of \gls{TFQKD} to realistic network scenarios.

\section{Conclusion}\label{sec:conclusions}

\gls{TFQKD} is one of the most promising candidate solutions for extending the range of real implementation of QKD in fiber. Here, we have discussed the impact of the dominant noise sources
in \gls{TFQKD} protocols when implemented in real-world conditions, providing a significant contribution toward their in-field use. We provided an open and unified framework for the modeling of major noise sources and the estimation of key rates, and we specifically addressed the phase noise of the photon sources and connecting fibers, showing how implementation aspects such as the quality of the used lasers, the adopted topology, the fiber length and unbalance in the two arms play a role in the final key rate and duty cycle. 
Interestingly, we observed that both the \gls{CAL} and \gls{SNS} protocols are impacted by phase noise in a similar way, although the relevant parameters enter the process via different mechanisms. We also highlighted the role of detector performance in significantly affecting the maximum achievable distance.
We showed as well how the overall key rate can be improved by a factor $\gtrsim 2$ using narrow-linewidth lasers and phase-control techniques as those developed to compare remote optical clocks on continental scales. With best-in-class but realistically deployable setups, we show that distances up to $~\SI{500}{\km}$ can be considered.
Synergy with the concurrent development of high-precision time/frequency distribution services \cite{clonets} is thus advisable, to lower the cost of deployment and achieve optimal usage of \gls{TFQKD} equipment.
Ultrastable lasers are nowadays found on the market in plug-and-play, compact and portable setup, and the technology is rapidly evolving towards further integration and miniaturization. We envisage these to be fruitfully combined with advanced phase-stabilization procedures~\cite{zhou2023}, towards an efficient and flexible \gls{TFQKD} implementation strategy on existing networks. 
These approaches would also allow to relax the constrains to the fiber network operators, hence facilitating the adoption of \gls{TFQKD} on a larger and operative scale. A critical application will be the establishment of long-haul links in the upcoming European Quantum Communication Infrastructure \cite{EuroQCI}, aiming at securely connecting distant quantum metropolitan area networks.
Very recently, new \gls{MDIQKD} protocols such as mode-pairing \gls{QKD} were proposed, that although first-order insensitive on laser and fiber noise, may still benefit from their active stabilization \cite{zhou2023}.
Also, we prospect that our considerations can be useful in the implementation of continuous-variable \gls{QKD}~\cite{Pirandola_Highratemeasurementdeviceindependentquantum_2015,Dias_Quantumrepeatercontinuousvariable_2020}, where similar challenges are encountered.

Proper handling of the phase noise and practical constraints of a given real-world network enables to consistently predict the expected key rate of a \gls{TFQKD} link and choose the optimal protocol, layout design and operating parameters depending on the network topology, available infrastructure and target performance. The code for the reproduction of the figures of this study, and for the estimation of key rates with varying setup and protocol parameters, is openly available at Ref.~\cite{bertaina_zenodo_2024}.

\begin{acknowledgments}
The results presented in this article have been achieved in the context of the following projects: QUID (QUantum Italy Deployment) and EQUO (European QUantum ecOsystems) which are funded by the European Commission in the Digital Europe Programme under the grant agreements number 101091408 and 101091561; 
Qu-Test, which has received funding from the European Union’s Horizon Europe under the Grant Agreement number 101113901; 
project ARS01\_00734-QUANCOM (European structural and investment funds MUR-PON Ricerca \& Innovazione 2014-2020);
EMPIR 19NRM06 METISQ, that received funding from the EMPIR program cofinanced by the Participating States and from the European Union Horizon 2020 research and innovation program; 
NATO Grant SPS G6026.
\end{acknowledgments}

\appendix

\section{Sending or not sending protocol}\label{app:SNSprotocol}

In this Appendix the \gls{SNS} protocol is discussed, starting with a description of the protocol, the estimation of the secret-key rate and then a highlight on how to include errors coming from phase instability in the channel. After its first proposal \cite{Wang_Twinfieldquantumkey_2018}, it attracted significant interest, with several works improving its security in practical cases (see, e.g., \cite{Yu2019}), increasing the achievable range \cite{Xu_Sendingornotsendingtwinfieldquantum_2020} and comparing it with other \gls{TFQKD} solutions \cite{Minder_Experimentalquantumkey_2019,Pittaluga2021}.

It can be partitioned in the following steps:
\begin{itemize}
 \item at each time slot, the parties commit to a Signal window with probability $p_\text{Z}$ or to a Decoy window with probability $p_\text{X}=1-p_\text{Z}$.
 \item if Alice (Bob) chooses Signal, with probability $\epsilon$, she (he) decides \textit{sending} and fixes a bit value 1 (0). With probability $(1 - \epsilon)$ she (he) decides \textit{not-sending} and fixes a bit value 0 (1);
 \item if \textit{sending} was chosen, \emph{they} send a phase-randomized weak coherent state $\ket{\sqrt{\mu_\text{Z}}}\exp(i\phi')$, with intensity $\mu_\text{Z}$ and phase $\phi'$ (never disclosed);
 \item following the decision of \textit{not-sending}, \emph{they} send out the vacuum state (or, more generally, a phase-randomized coherent state with very small intensity $\mu_0$). Notice that \emph{sending} or \emph{not sending} determines the bit value, not the intensity, phase or photon number;
 \item if \emph{they} chose Decoy, \emph{they} send out a phase-randomized coherent state with intensity randomly chosen from a predetermined set $\ket{\sqrt{\mu_k}}\exp(i\phi'), k=1,2,3...$. Note that the phase values in the decoy windows will be disclosed after the end of the whole transmission session, in order to reconcile the phase slices and estimate the phase error rate;
 \item Afterwards, \emph{they} classify the time windows in the following way:
 \item Z window: \emph{they} both chose Signal;
 \item $\tilde{Z}$ window: Z window in which only one sends;
 \item $\tilde{Z}_1$ window: $\tilde{Z}$ window in which a single-photon state is sent. This may contribute to the key rate;
 \item $X_k$ window: \emph{they} both chose Decoy, and the same decoy intensity $\mu_k$;
 \item effective window: Charlie announces only one detector click. Only these cases may contribute to building the final key. Double-click and zero-click events are discarded and they therefore contribute to the bit-flip \gls{QBER};
 \item Key distillation starts by the declaration by Charlie of the $n_\text{t}$ effective Z windows;
 \item \emph{they} publicly choose a small sample of effective Z windows, which will have to be discarded (asymptotically this is negligible), for estimating the bit-flip error rate $E_\text{Z}=(n_{\text{NN}}+n_{\text{SS}})/n_\text{t}$. For this sample, \emph{they} indeed disclose whether \emph{they} both chose sending (rate $n_{\text{SS}}/n_\text{t}$) or both not-sending (rate $n_{\text{NN}}/n_\text{t}$);
 \item estimate number of untagged bits: only effective $\tilde{Z}$ windows. Indeed multiphoton signals must be considered tagged (attacked by Eve). Their number cannot be measured. Their lower bound $\underaccent{\bar}{n}_1$, and the upper bound $\bar{e}_1^{\text{ph}}$ of their phase error, can be estimated using decoy states, in particular using effective $X_k$ windows (see Appendix~\ref{app:decoy}).
\end{itemize}

The secret key per transmitted signal with unity duty cycle can be estimated with the following expression \cite{Xu_Sendingornotsendingtwinfieldquantum_2020} 
\begin{equation}
\label{SNSrate1}
\underaccent{\bar}{R} = p_\text{Z}^2 \left[\underaccent{\bar}{n}_1 \left(1-H_2(\bar{e}_1^{\text{ph}})\right)- f_{\text{EC}} n_\text{t} H_2(E_\text{Z})\right] \ .
\end{equation}
Asymptotically we let the sifting factor $p_\text{Z}^2=1$, however notice that the decoy measurements might be inefficient with large enough phase errors or losses, or short keys, so that it could be unrealistic to fix $p_\text{X}\approx 0$.

In order to keep under control the bit-flip error rate, small values of the sending probability $\epsilon$ must be chosen (a few percent). Error rejection techniques \cite{Xu_Sendingornotsendingtwinfieldquantum_2020} can be applied before the parameter estimation stage to reduce the bit-flip errors, allowing the use of larger values of $\epsilon$. The steps can be organized as follows: 
\begin{itemize}
 \item First of all, one can take into account in the key rate expression that the bit-flip error rates for bits 0 and 1 are intrinsically different in this protocol.
 \item Afterwards, the parties perform synchronized random pairing of their raw key bits. Then, they compare the parity of the pairs, discarding pairs with different parity and keeping the first bit of pairs with the same parity. The effect is a rejection of a fraction of bit-flip errors, at the cost of a cut in the length of the raw key.
 \item By scrutinizing the residual bit-flip error rate of the survived bits, it turns out that it is still high for even-parity pairs. One can keep just the odd-parity pairs, the number of which will be on average $N_{\text{odd}}=N_0 N_1/(N_0+N_1)$, where $N_0$ ($N_1$) represent the number of 0s (1s) in the raw key string of Bob.
 \item \gls{SNS-AOPP}: the last evolution consists in substituting the random pairing with actively pairing the bits in odd-parity pairs. In this case, the number of odd-parity pairs will increase to $N_{\text{odd}}^{\text{AOPP}}=\min(N_1,N_0)$. Also in this case, only the first bit in each pair is kept. The final key rate can be estimated as Eq.~\ref{SNSrate2},
 where $n_\text{t}$ is the length of the string after AOPP, of which $n'_1$ are untagged, while $\bar{e}_1^{' ph}$ and $E'_\text{Z}$ are the phase and bit-flip error rate after AOPP. Complete expressions can be found in \cite{Xu_Sendingornotsendingtwinfieldquantum_2020}.
\end{itemize}

\section{Curty-Azuma-Lo protocol}\label{app:CALprotocol}

The \gls{CAL} protocol was proposed in 2019 by M. Curty, K. Azuma and H.-K. Lo in \cite{CurtyAzumaLo2019} and a proof-of-principle experimental demonstration can be found in \cite{Zhong2019}. The protocol relies on the pre-selection of a global phase and is conceptually very simple, consisting in the following steps:
\begin{itemize}
 \item First of all, Alice (Bob) chooses with probability $p_\text{X}$ the X basis (key generation) and with probability $p_\text{Z}=1-p_\text{X}$ the Z basis (control). In the time slots in which her (his) choice was the X basis, she (he) draws a random bit $b_\text{A}$ ($b_\text{B}$). Then, she (he) prepares an optical pulse $a$ ($b$) in the coherent state $\ket{\zeta}_{a(b)}$ for $b_\text{A}=0$ ($b_\text{B}=0$) or $\ket{-\zeta}_{a(b)}$ for $b_\text{A}=1$ ($b_\text{B}=1$). In the time slots in which her (his) choice is the Z basis, she (he) prepares an optical pulse $a$ ($b$) in a phase-randomized coherent state $\hat{\rho}_{a,\beta_\text{A}}$ ($\hat{\rho}_{B,\beta_\text{B}}$) where the amplitude $\beta_\text{A}$ ($\beta_\text{B}$) is chosen from a set $S=\{\beta_i\}_i$ of real non-negative numbers $\beta_i\geq 0$, according to a probability distribution $p_{\beta_\text{A}}$ ($p_{\beta_\text{B}}$).
 \item Alice and Bob transmit the optical pulses $a$ and $b$ over channels with transmittance $\sqrt{\eta}$ towards the middle node C and synchronize their arrival.
 \item Node C interferes the incoming optical pulses $a$ and $b$ on a 50:50 beamsplitter. The output ports are coupled to two threshold detectors, $D_\text{c}$ and $D_\text{d}$, associated respectively to constructive and destructive interference.
 \item C announces publicly the measurement outcomes $k_\text{c}$ and $k_\text{d}$ corresponding to detectors $D_\text{c}$ and $D_\text{d}$. A click event is indicated by $k_i=0$ and a no-click event by $k_i=1$, with $i=\text{c},\text{d}$.
 \item The raw key is generated by Alice and Bob concatenating the bits $b_\text{A}$ and $b_\text{B}$ ($b_\text{A}$ and $b_\text{B} \oplus 1$) when node C announces $k_\text{c}=1$ and $k_\text{d}=0$ ($k_\text{c}=0$ and $k_\text{d}=1$) and Alice and Bob chose the $X$ basis.
\end{itemize}

The protocol requires a common phase reference between Alice and Bob for the key generation basis. Local phase randomization is applied in the $Z$ basis, allowing the application of the decoy-state technique to infer the contribution of vacuum, single-photon and multi-photon events.
For the security proof, the authors invoke a "complementarity" relation between the phase and the photon number of a bosonic mode. The details can be found in \cite{CurtyAzumaLo2019}.

The final secret key per time slot can be lower bounded by the following expression, summing the contribution from the single-click events ($k_\text{c}=1 , k_\text{d}=0$) and ($k_\text{c}=0$ , $k_\text{d}=1$)
\begin{equation}
 \underaccent{\bar}{R}_\text{X}= \underaccent{\bar}{R}_{X,10} + \underaccent{\bar}{R}_{X,01} \ ,
\end{equation}
where
\begin{align}
 \label{CALrate1}
 &\underaccent{\bar}{R}_{X,k_\text{c} k_\text{d}}=p_\text{XX}(k_\text{c},k_\text{d}) \\ 
 &[ 1-f_{\text{EC}} H_2(e_{X,k_\text{c} k_\text{d}})-H_2(\text{min}\{
1/2,\bar{e}_{Z,k_\text{c} k_\text{d}} \}) ] \ . \nonumber
\end{align}
In the expression above, $p_\text{XX}(k_\text{c},k_\text{d})$ represents the total gain when Alice and Bob choose the $X$ basis, $e_{X,k_\text{c} k_\text{d}}$ is the bit error rate in the X basis, while $\bar{e}_{Z,k_\text{c} k_\text{d}}$ is the upper bound on the phase error rate. The estimation of these quantities is detailed in the following paragraphs.

The total gain for the generation events can be expressed as
\begin{align}
 \label{CALgain}
 &p_\text{XX}(k_\text{c},k_\text{d})=\frac{1}{4} \sum_{b_\text{A},b_\text{B}=0,1} p_\text{XX}(k_\text{c},k_\text{d}|b_\text{A}, b_\text{B}) = \\ 
 &= \frac{1}{2}(1-p_\text{d}) \left( e^{-\gamma \Omega(\sigma_\varphi,\theta)} + e^{\gamma \Omega(\sigma_\varphi,\theta)} \right) e^{-\gamma} - (1-p_\text{d})^2 e^{-2\gamma} \ . \nonumber
\end{align}
The second expression is obtained by modeling the channel for simulations, see supplementary information of \cite{CurtyAzumaLo2019} for details. The model consists in a loss $\sqrt{\eta}$, a phase mismatch $\sigma_\varphi$ and a polarization mismatch $\theta$, giving rise to the parameters $\gamma=\sqrt{\eta} \mu_\zeta$ (with $\mu_\zeta=|\zeta|^2$ the intensity of the signal states) and $\Omega=\cos{\sigma_\varphi} \cos{\theta}$.

The bit error rate can also be estimated by the channel model as
\begin{equation}
 \label{CALex}
 e_{X,k_\text{c} k_\text{d}}= \frac{e^{-\gamma \Omega(\sigma_\varphi,\theta)}-(1-p_\text{d})e^{-\gamma}}{e^{-\gamma \Omega(\sigma_\varphi, \theta)}+e^{\gamma \Omega(\sigma_\varphi, \theta)}-2(1-p_\text{d})e^{-\gamma}} \ .
\end{equation}

The phase error rate requires a more involved analysis. Following Eqs.~10 to 15 of \cite{CurtyAzumaLo2019} and its supplementary material, one can obtain the following expression for the upper bound on the error in the $Z$ basis
\begin{align}
 \label{CALez}
 &\bar{e}_{Z,k_\text{c} k_\text{d}} \leq \frac{1}{p_\text{XX}(k_\text{c}, k_\text{d})} \sum_{j=0,1} \Bigg[ \sum_{(m_\text{A}, m_\text{B}) \in \mathcal{S}_j} c_{2m_\text{A}+j}^{(j)} c_{2m_\text{B}+j}^{(j)} \\ \nonumber
 &\times \sqrt{\bar{p}_\text{ZZ}(k_\text{c}, k_\text{d}|2m_\text{A}+j,2m_\text{B}+j)}+\Delta_j \Bigg]^2 \ .
\end{align}
Only the $\bar{p}_\text{ZZ}$ gains for low number of photons (defined in the set $\mathcal{S}_j$), while the other probabilities are trivially upper-bounded by 1 and are included in the term $\Delta_j$.
In \cite{CurtyAzumaLo2019} a numerical method to estimate the $\bar{p}_\text{ZZ}$ gains for a finite number of decoy intensities is reported. In \cite{Grasselli2019}, instead, these quantities are estimated analytically for two, three and four decoy intensities. Since one can show that realistic implementations with 3 or 4 decoy intensities are almost optimal, in this work the gains are analytically estimated assuming an infinite number of decoy intensities, following the supplementary material of \cite{CurtyAzumaLo2019}. Similarly to the simulations in \cite{CurtyAzumaLo2019}, the sets are chosen as $\mathcal{S}_0=\{(0,0),(0,1),(1,0),(1,1) \}$ and $\mathcal{S}_1={(0,0)}$ and the bounds may be improved by adding more terms in the estimation. It turns out that the phase error rate is independent of the phase mismatch, while it has an important effect on the bit error rate. 
To keep the phase error rate low enough, small values of the signal intensity must be chosen, around $0.02$. In the \gls{SNS} protocol, on the other hand, small values of the sending probability are chosen to lower the error rate, leading to comparable effects on the key rate. 

\section{Decoy state expressions}\label{app:decoy}

For completeness, we report here the expressions for the error estimates in the three-decoy-state approach\cite{Lo_DecoyStateQuantum_2005,Ma_Practicaldecoystate_2005,Tamaki_Decoystatequantumkey_2016}, that we used both in the phase-encoded \gls{BB84} calculations and for the phase error in the \gls{SNS} protocol.

The gain for each laser intensity $\mu=u,v,w$ and effective transmission $\hat{\eta}$ is:
\begin{equation}
 Q_\mu=1- (1-p_{\text{DC}})e^{-\mu\hat{\eta}}\,.
\end{equation}
The corresponding total QBER is modeled as
\begin{equation}
E_\mu=\left[\frac{p_{\text{DC}}}{2} + \left(e_\theta+e_\varphi-\frac{p_{\text{DC}}}{2}\right) e^{-\mu\hat{\eta}}\right]/Q_\mu
\end{equation}
where $e_\theta$ and $e_\varphi$ are the optical and the phase noise errors, respectively, as defined in the main text. For the phase-encoded \gls{BB84} protocol, which has only a single channel of length $L_\text{A}+L_\text{B}$ along which interfering photons are separated only by few ns, we assume $e_\varphi=0$.

Assuming that intensity $\mu=u$ is matched to the relevant signal intensity and that $u>v>w$, then the lower bounds for the zero and single-photon yield are given by
\begin{align}
\underaccent{\bar}{Y}_0 &= \frac{v Q_w e^w-w Q_v e^v}{v-w}\\
\underaccent{\bar}{Y}_1 &= \frac{u^2(Q_v e^v-Q_w e^w)-(v^2-w^2)(Q_u e^u-\underaccent{\bar}{Y}_0)}{u(u-v-w)(v-w)}\;,
\end{align}
so that the single-photon lower bound for the gain and upper bound for the phase error are estimated by
\begin{align}
\underaccent{\bar}{Q}_1 &= \underaccent{\bar}{Y}_1 u e^{-u} \\
\bar{e}_1^{\text{ph}} &=\frac{E_v Q_v e^v-E_w Q_w e^w}{(v-w)\underaccent{\bar}{Y}_1}
\end{align}

In case of the phase-encoded \gls{BB84} model, the key rate expression that we use is 
\begin{equation}
 R = d \left[\underaccent{\bar}{Q}_1 \left(1-H_2(\bar{e}_1^{\text{ph}})\right)- f_{\text{EC}} Q_u H_2(E_u)\right] \ ,
\end{equation}
with duty cycle asymptotically set to $d=1$ in the efficient imbalanced basis selection setting.

\section{Parameters of key rate simulations}\label{app:simparameters}

\begin{table}[!tbh]
\renewcommand{\arraystretch}{1.5}
 \centering
 \begin{tabular}{c|c|c|c|c|c|c|c}
 $\alpha$ & $\nu_\text{s}$ & $\tau_{\text{PS}}$ & $e_\theta$ & $P_{\text{DC}}^\text{SNSPD}$ & $\eta_{\text{D}}^\text{SNSPD}$ & $P_{\text{DC}}^\text{SPAD}$ & $\eta_{\text{D}}^\text{SPAD}$ \\
 \SI{0.2}{\decibel\per\km} & \SI{1}{\GHz} & \SI{1}{\milli\second}& $0.02$ & \SI{10}{\Hz} & $0.9$ & \SI{50}{\Hz} & $0.25$ \\
 \hline
 $u$ & $v$ & $w$ & $\mu_\text{Z}$ & $\mu_0$ & $\mu_\zeta$ & $\epsilon$& $f_{\text{EC}}$ \\
 $0.4$ & $0.16$ & $10^{-5}$ & $0.2$ & $5\times10^{-6}$ & $0.018$ & $0.25$ & $1.15$\\
 \end{tabular}
 \caption{Parameters of key rate simulation.}
 \label{tab:simulationparameters}
\end{table}

Common parameters employed in the simulations are reported in Table~\ref{tab:simulationparameters} and discussed here. A representative attenuation coefficient $\alpha=\SI{0.2}{\decibel\per\km}$ is chosen, considering that the typical attenuation in real-field can exceed $\SI{0.25}{\decibel\per\km}$, while new-generation laboratory fibers reach $\SI{0.16}{\decibel\per\km}$~\cite{Minder_Experimentalquantumkey_2019,Liu_ExperimentalTwinFieldQuantum_2023}. As a best-in-class commercial \gls{SNSPD}, we consider dark count rate $P_{\text{DC}}^\text{SNSPD}=\SI{10}{\Hz}$, corresponding to dark counts per signal $p_{\text{DC}}^\text{SNSPD}=P_{\text{DC}}^\text{SNSPD}/\nu_\text{s}=10^{-8}$, with efficiency $\eta_{\text{D}}^\text{SNSPD}=\SI{90}{\percent}$ and nominal source clock rate $\nu_\text{s}=\SI{1}{\GHz}$.
As a more common best-in-class commercial \gls{SPAD}, we consider dark count rate $P_{\text{DC}}^\text{SPAD}=\SI{50}{\Hz}$, corresponding to dark counts per signal $p_{\text{DC}}^\text{SPAD}=5\times10^{-8}$, with efficiency $\eta_{\text{D}}^\text{SPAD}=\SI{25}{\percent}$.
The total intensities for the three decoy states $u, v, w$ used in the \gls{SNS-AOPP} and phase-encoded \gls{BB84} protocols are taken from Ref.~\cite{Minder_Experimentalquantumkey_2019}. Intensities for the \emph{sending} and \emph{not sending} choices in the \gls{SNS-AOPP} protocol are set to $\mu_\text{Z}=u/2$ and $\mu_0=w/2$, respectively, while intensity for the Alice signal in \gls{BB84} corresponds to $u$, and signal intensity in the CAL protocol is set to the value optimized in \cite{Zhong2019}. 

\section{Derivation of the common-laser phase noise spectrum}
\label{app:common-spectrum}

Following the scheme shown in Figure~\ref{fig:tfqkdsetup}b, let us assume that the instantaneous phase of the reference laser in Charlie is $\varphi_\text{l,C}$. While traveling to Alice/Bob, the signal acquires additional phase $\varphi_\text{F,X}$. Here and in the following we adopt a compact notation in which $\varphi_\text{F,X}(t_\text{out})$ identifies the \textit{integrated} phase of a fiber with length $L_\text{X}$ accumulated during the whole journey from the moment radiation enters in it ($t_\text{out}-nL_\text{X}/c$) till the moment it exits $t_\text{out}$. Photon sources in Alice and Bob are phase-locked to incoming light and have therefore instantaneous phase $\varphi_\text{l,X}(t)=\varphi_\text{l,C}(t-nL_\text{X}/c) + \varphi_\text{F,X}(t)$. This is a replica of the original reference laser phase, with additive noise due to propagation in the fiber. In turns, these photons are sent to Charlie, acquiring further phase due to backward-trip in the quantum fiber. 
Assuming the noise of the auxiliary and quantum fibers to be highly correlated (this is justified as they are housed into the same optical cable), the relative phase of interfering photons in Charlie at a time $t$ is thus rewritten as:
\begin{equation}
\label{eq:lasphi}
\begin{split}
\Delta\varphi(t) &= 
\varphi_\text{l,C}(t-2nL_\text{A}/c)+\varphi_\text{F,A}(t-nL_\text{A}/c)+
 \varphi_\text{F,A}(t) \\
 & - \varphi_\text{l,C}(t-2nL_\text{B}/c)-\varphi_\text{F,B}(t-nL_\text{B}/c) -\varphi_\text{F,B}(t)
\end{split}
\end{equation}
Under the assumption that the fiber deformations change on timescales much longer than the light round-trip time, $\varphi_\text{F,X}(t)\approx \varphi_\text{F,X}(t-nL_\text{X}/c)$ and these two terms add up coherently. 
Computing the autocorrelation function of the various terms of Eq.~\ref{eq:lasphi} and the corresponding Fourier transforms, and using the property that the Fourier transform $\mathcal{F} [y(t+\Delta)]=e^{2\pi f i \Delta}\mathcal{F}[y(t)]$, Eq.~\ref{eq:noise_1laser} follows. 

In the case a sensing laser is used to 
determine the instantaneous fiber phase variations, the corresponding interferometric error signal upon interference in Charlie can be computed adopting the same reasoning and takes the same form as Eq.~\ref{eq:lasphi}, with the subscript $s$ instead of $r$. It follows that, if $\varphi_\text{r} \approx \varphi_\text{s}$, i.e. the two lasers are phase-coherent, the error signal derived by interfering the sensing laser can be exploited to cancel residual noise of the common reference laser in addition to the fiber noise, and further improve the phase stability. 

\section{Models for the laser noise}
\label{app:laser_phase_noise}

In general, the noise of standard diode lasers used in frequency dissemination follows a law of the type:
\begin{equation}
\label{eq:freelas}
S_\text{l, free}(f) =\frac{r_{3}}{f^3}+\frac{r_{2}}{f^2} \left( \frac{f_\text{c}}{f+f_\text{c}}\right)^2
\end{equation}
where $r_{3}$ and $r_{2}$ depend on the laser technology, and the cutoff frequency $f_\text{c}$ is related to the modulation (control) bandwidth of the laser. 
The linewidth of these lasers is typically of the order of \SI{1}{\kHz} to \SI{100}{\kHz} and the coherence time is $<$\SI{100}{\micro\s}, even though performances of commercially-available solutions are continuously improving \cite{bouchand2017,liang2023}.
Narrow-linewidth lasers can grant superior phase-coherence between successive realignments. They can be realized in several ways, e.g. nanofabrication \cite{loh2020,lee2013}, delay-line stabilization \cite{kefelian2009} or external high-finesse cavity stabilization \cite{herbers2022,matei2017,kelleher2023}. We provide coefficients for the latter approach as it is the one with best performances today. The interested reader can refer to the literature for the optimal compromise in terms of size, weight and power vs performances.

\begin{table*}[ptb]
 \centering
 \setlength{\extrarowheight}{2pt}
 \begin{tabular}{m{0.1\textwidth}|M{0.32\textwidth}|M{0.55\textwidth}} 
 {Laser (free)} &{$S_\text{l,free}(f) =\frac{r_{3}}{f^3}+\frac{r_{2}}{f^2} \left( \frac{f_\text{c}}{f+f_\text{c}}\right)^2$} &
 \begin{tabular}{M{0.27\columnwidth}M{0.27\columnwidth}M{0.27\columnwidth}M{0.27\columnwidth}}
 $r_{3}$ & $r_{2}$ & $f_\text{c}$ &\\
 \SI{3e6}{\radian\squared \Hz\squared} & \SI{3e2}{\radian\squared \Hz} & \SI{2}{\MHz} &
 \end{tabular}
 \\
 \hline
 Laser (stable) & $S_\text{l,stab}(f)=S_\text{cavity}(f)+\left |\frac{1}{1+G(f)} \right|^2 S_\text{l,free}(f) $ &
 \begin{tabular}{M{0.27\columnwidth}M{0.27\columnwidth}M{0.27\columnwidth}M{0.27\columnwidth}}
 &&\\&&
 \end{tabular}
 \\
 \hline
 {Cavity} &{$S_\text{cavity}(f) =\frac{C_{4}}{f^4}+\frac{C_{3}}{f^3}+\frac{C_{2}}{f^2}$} &
 \begin{tabular}{M{0.27\columnwidth}M{0.27\columnwidth}M{0.27\columnwidth}M{0.27\columnwidth}}
 $C_{4}$ & $C_{3}$ & $C_{2}$ &\\
 \SI{0.5}{\radian\squared\Hz\cubed} & \SI{0}{\radian\squared \Hz\squared} & \SI{2e-3}{\radian\squared \Hz} &
 \end{tabular}
 \\
\hline
 Loop & $G(f) = G_0\frac{1}{(2\pi i f)^2}\frac{if+ B\gamma}{if+ B\delta}$ &
 \begin{tabular}{M{0.27\columnwidth}M{0.27\columnwidth}M{0.27\columnwidth}M{0.27\columnwidth}}
 $B$ & $\gamma$ & $\delta$ & $G_0$\\
 \SI{300}{\kHz} & $0.1$ & $10$ & \SI{3.55e13}{\Hz\squared}
 \end{tabular}
 \\
\hline
 {Fiber (free)} &{$S_\text{F}(f,L) = \frac{l L}{f^2}
 \left(\frac{f_\text{c}'}{f+f_\text{c}'}\right) ^2 $} &
 \begin{tabular}{M{0.27\columnwidth}M{0.27\columnwidth}M{0.27\columnwidth}M{0.27\columnwidth}}
 $l$ & $f_\text{c}'$ & &\\
 \SI{44}{\radian\squared \Hz\per\km} & \SI{100}{\Hz} &
 \end{tabular}
 \\
\hline
 Fiber (stable) & $S_\text{F,s}(f,L) = \frac{(\lambda_\text{s}-\lambda_\text{q})^2}{\lambda_\text{s}^2} \frac{l L}{f^2}+s_0 \left (\frac{f_\text{c}''}{f+f_\text{c}''} \right )^2$ &
 \begin{tabular}{M{0.27\columnwidth}M{0.27\columnwidth}M{0.27\columnwidth}M{0.27\columnwidth}}
 $s_0$ & $f_\text{c}''$ & $\lambda_\text{s}$ & $\lambda_\text{q}$\\
 \SI{1e-8}{\radian\squared\per\Hz} & \SI{200}{\kHz} & \SI{1543.33}{\nm} & \SI{1542.14}{\nm}
 \end{tabular}
 \end{tabular} 
 \caption{Recap of analytical models for the various terms needed to evaluate the phase jitter and coefficients extrapolated from our experimental data.}
 \label{tab:coeffs}
\end{table*}

The noise of a cavity-stabilized laser 
$S_\text{l,stab}(f)$ depends on the local cavity noise $S_\text{cavity}(f)$ at low Fourier frequencies, and on the intrinsic noise of the used laser source $S_\text{l,free}(f)$ (Eq.~\ref{eq:freelas}) at high Fourier frequencies, via the gain function $G(f)$ that regulates the control loop response: 
\begin{equation}
\label{eq:stablas}
 S_\text{l,stab}(f)=S_\text{cavity}(f)+\left |\frac{1}{1+G(f)} \right|^2 S_\text{l,free}(f)
\end{equation}
The cavity noise is usually parametrized by:
\begin{equation}
 \label{eq:cavnoise}
 S_\text{cavity}(f) =\frac{C_{4}}{f^4}+\frac{C_{3}}{f^3}+\frac{C_{2}}{f^2}
\end{equation}
with coefficients that depend on the cavity material, geometry and passive isolation, and on technical noise \cite{matei2017,herbers2022,kelleher2023}. Typical coefficients for a compact, portable cavity system, that appears suited for the considered application, are reported in Table \ref{tab:coeffs}.

The loop function model follows general concepts of control theory, and includes considerations on the bandwidth allowed by all sub-system. The overall loop function can be described by a complex function in the Laplace space:
\begin{equation}
\label{eq:G}
G(f) = G_0\frac{1}{(2\pi i f)^2}\frac{if+ B\gamma}{if+ B\delta}
\end{equation}
which includes a second-order integrator to provide high gain at low frequencies, a single integrator stage emerging at a corner frequency $B\gamma$, 
with $B$ the loop bandwidth and $\gamma < 1$, and is ultimately limited by the finite response of the system, featuring at least one pole at frequency $B\delta$, with $\delta > 1$. The parameters $\gamma$, $\delta$ determine the exact positions of the knees in the loop response (zero and pole respectively) relatively to the loop bandwidth, $G_0 = (2\pi B)^2 (1 + \delta)/(1+\gamma)$. All of these terms are fine-tuned empirically to maximize the noise rejection and adapt to possible poles present in the subsystems transfer function, but good design values are $\gamma\approx 0.1$ and $\delta\approx 10$.

\section{Derivation of model coefficients}
\label{app:noise_coefficients}

\begin{figure}[!tbp]
 \centering
 \includegraphics[width=\columnwidth]{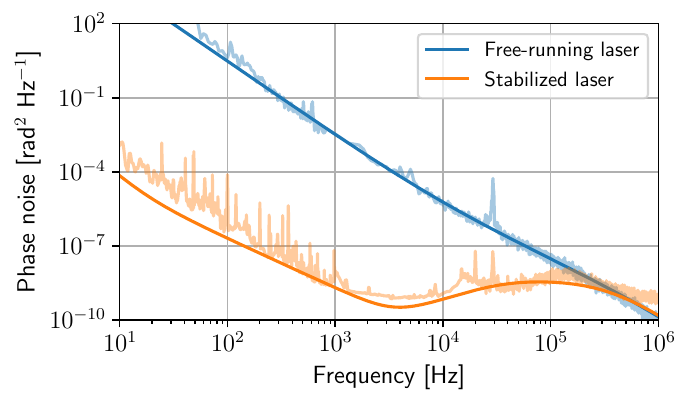}
 \caption{Measured (lighter) and modeled (darker) values for a free-running (blue) and cavity-stabilized (orange) diode laser noise.}
 \label{fig:lasernoise}
\end{figure}

\begin{figure}[!tbp]
 \centering
\includegraphics[width=\columnwidth]{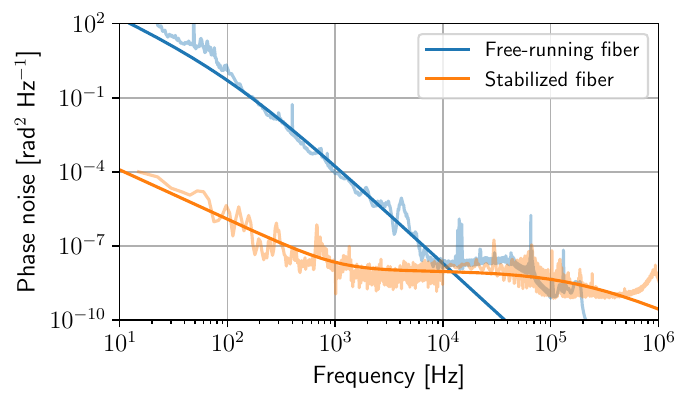}
 \caption{Measured and modeled noise of a \SI{114}{\km} long free-running (blue) and stabilized (orange) fiber, traveled in a double-pass~\cite{Clivati_Coherentphasetransfer_2022}. To account for the double-pass, the instance of the fiber noise model is multiplied by a factor of 4.}
 \label{fig:fibnoise}
\end{figure}

Figure~\ref{fig:lasernoise} (blue) shows the experimentally measured power spectral density of a $<$\SI{10}{\kHz}-linewidth planar waveguide extended-cavity diode laser in a butterfly package (PLANEX by RIO Inc., see also \cite{clivati2011, numata2009,bouchand2017,liang2023} for other laser types), and the noise of the same laser when stabilized to an external \SI{5}{\cm}-long Fabry-Perot cavity with Finesse exceeding 100'000 (orange). Darker shades represent instances of the respective models according to Eqs.~\ref{eq:freelas} and \ref{eq:stablas} for the coefficients shown in Table \ref{tab:coeffs}.
The spur observed at about \SI{30}{\kHz} both on the free-running and stabilized laser (in the latter case, reduced by a factor corresponding to the stabilization loop efficiency at this frequency) is not considered by the model and is attributed to an electrical disturbance on our diode laser current driver. Especially on the stabilized laser below \SI{1}{\kHz}, noise peaks are found at specific frequencies: they are due to residual acoustic and seismic solicitations of the resonator and do not significantly affect the results.

Deriving a unique estimate for the fiber noise on a generic layout is more complicated, as the $l$ coefficient primarily depends on the environment where the fiber is housed \cite{clivati2018}.
In general, field noise levels exceed those of spooled fibers with equal length \cite{williams2008}, but up to a factor 10 variation is observed between the various installations (compare, e.g. values in \cite{williams2008,droste2013,Clivati_Coherentphasetransfer_2022,lopez2010,akatsuka2020,akatsuka2014,clivati2020,husman2021}).
As a reference, Figure~\ref{fig:fibnoise} (blue) shows the measured noise of a \SI{114}{\km} fiber traveled in a round-trip \cite{Clivati_Coherentphasetransfer_2022}. The fiber was deployed on an intercity haul running parallel a highway for the majority of its part. The corresponding model is obtained from Eq.~\ref{eq:freefibnoise} with a coefficient $l =\SI{44}{\radian\squared\Hz\per\km}$. In the figure, the modeled noise is multiplied by a factor of 4 to account for the fact that the fiber noise is measured in a round-trip.

Fiber noise in a stabilized condition is well explained by Eq.~\ref{eq:stabfibnoise} below \SI{1}{\kHz} of Fourier frequency, according to the fact that we stabilized the fiber at $\lambda_\text{s}=\SI{1543.33}{\nm}$ and observed the effect at $\lambda_\text{q}=\SI{1542.14}{\nm}$.
To explain the experimental observations at higher frequency, we also include a white phase noise detection floor of the form $S_\text{detection}(f) = s_0 \left [f_\text{c}''/(f+f_\text{c}'') \right ]^2$, with coefficient $s_0=\SI{1e-8}{\radian\squared\per\Hz}$ corresponding to a typical \gls{SNR} of \SI{80}{\decibel\radian\squared\per\Hz} for the sensing laser interference, upper-limited at a cutoff frequency of $f_\text{c}''=\SI{200}{\kHz}$. 

\section{Derivation of the integration time for scenarios shown in Sec.~\ref{sec:simulations}}
\label{app:qbertau}

\begin{figure*}[ptb]
 \centering
 \begin{tabular}{cc}
 \includegraphics[width=1\textwidth]{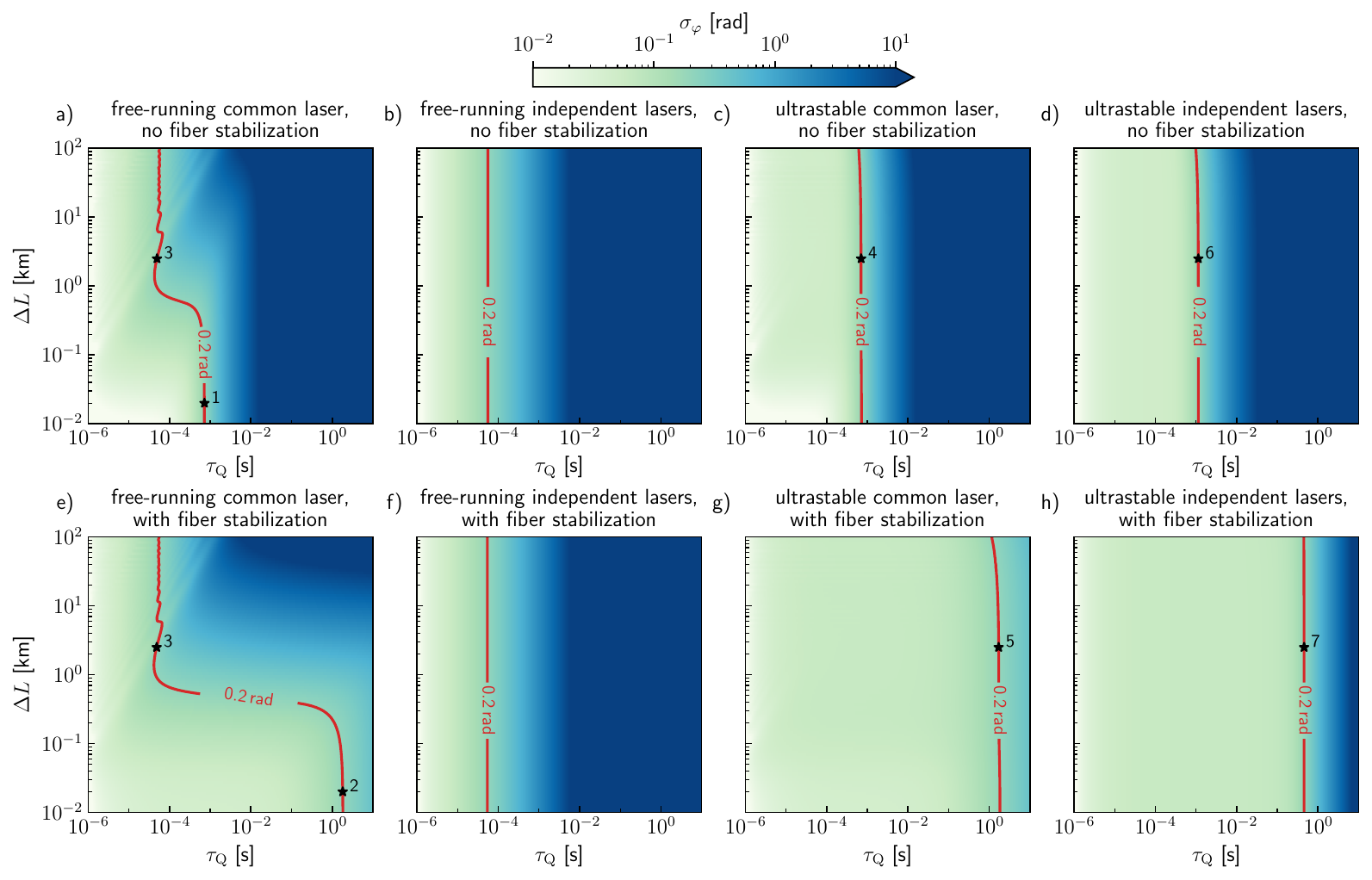}
 \end{tabular}
 \caption{Maps of the phase standard deviation $\sigma_\varphi$, calculated in the space of fiber length mismatch $\Delta L$ and integration time $\tau_\text{Q}$, for each possible combination of laser source configuration and fiber stabilization at fixed shorter arm $L_\text{B}=\SI{100}{\km}$. The isolines corresponding to characteristic $\sigma_\varphi$ values are also reported. The numbered stars represent the specific scenarios considered in Sec.~\ref{sec:simulations}, where representative values of $\Delta L$ are chosen.}
 \label{fig:sigmaunbalancemaps}
\end{figure*}

\begin{figure*}[ptb]
 \includegraphics[width=0.95\textwidth]{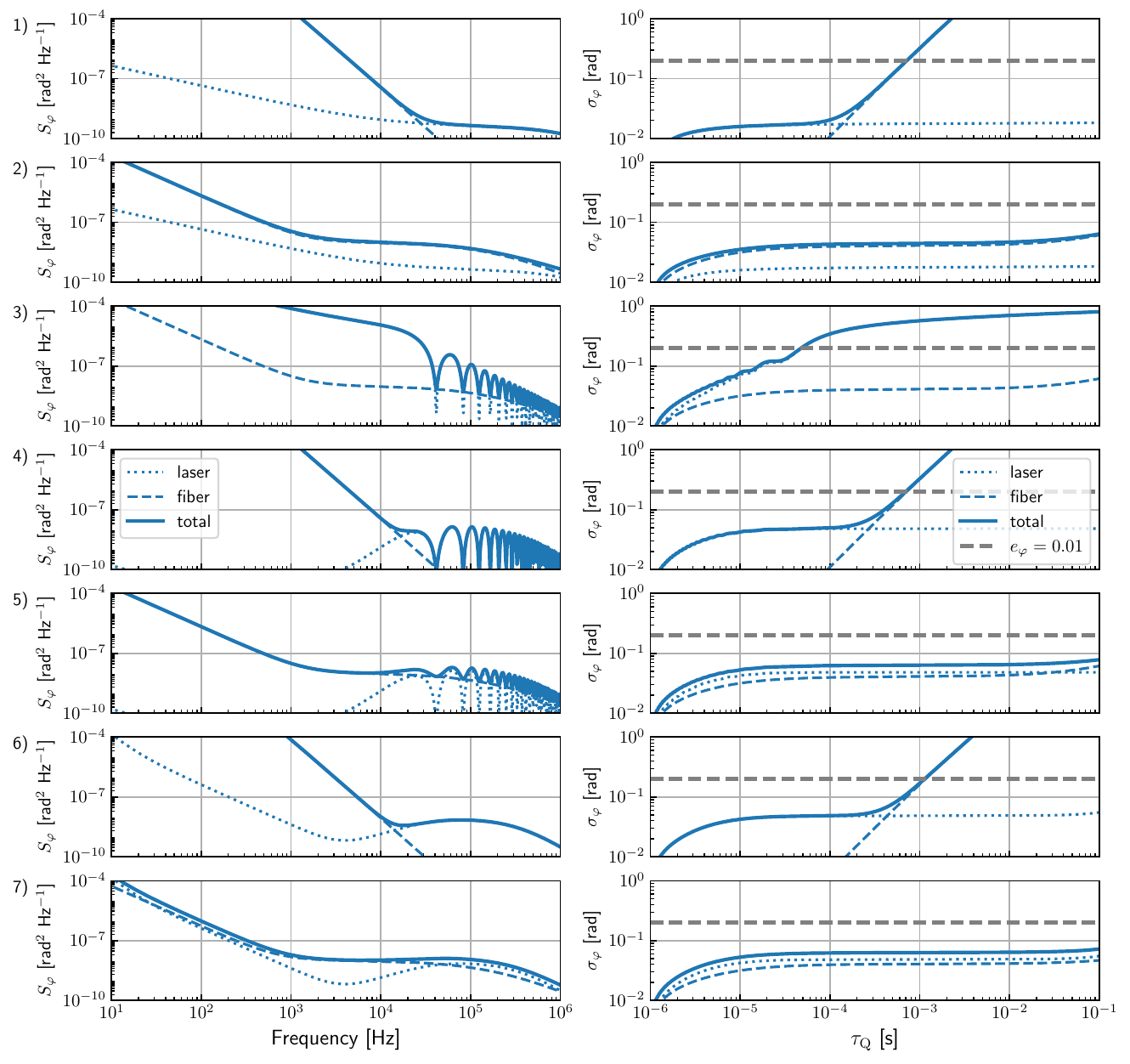}
 \caption{\label{fig:noisescenarios}Phase noise as a function of frequency (left) and phase variance as a function of the interrogation time (right) for the seven scenarios reported in Sec.~\ref{sec:simulations}, at fixed shorter arm $L_\text{B}=\SI{100}{\km}$. Noise expressions were derived from Eqs.~\ref{eq:noise_1laser} and \ref{eq:noise_2lasers}, together with Eqs.~\ref{eq:freelas},\ref{eq:stablas}, \ref{eq:freefibnoise},\ref{eq:stabfibnoise} for detailed expression of the lasers and fiber noise.}
\end{figure*}

Figure~\ref{fig:sigmaunbalancemaps} reports the full maps of $S_\varphi(f)$ and $\sigma_\varphi$, calculated for each combination of laser source configuration and fiber stabilization, which were discussed in Sec.~\ref{sec:simulations}. The contour lines that were plotted in Figure~\ref{fig:sigma_unbalance} were extracted from these maps, and the characteristic scenarios are represented as numbered points. In panels a and e, the strong apparent dependence of $\tau_\text{Q}$ on $\Delta L$ is related to the laser coherence length, as discussed for Figure~\ref{fig:sigma_unbalance}. Notice that, in Scenario 3, laser noise always dominates the total noise (See Figure~\ref{fig:noisescenarios}, panel 3 on the right).

Figure~\ref{fig:noisescenarios} shows the phase noise contributions of laser(s) and fibers, as well as their combined effect, for the seven scenarios of Sec.~\ref{sec:simulations}. Panels on the right indicate the corresponding phase standard error as a function of $\tau_\text{Q}$ derived from Eq.~\ref{eq:jit_spec} and the threshold corresponding to a phase-misalignment \gls{QBER} $e_\varphi=\SI{1}{\percent}$ as derived from Eq.~\ref{eq:qberfromvar}.
For evaluating the fiber noise we considered the shortest interferometer arm to be of length $L_\text{B}=$\SI{100}{\km}, and included length mismatches as indicated in Table~\ref{tab:scenarios}. 
We assume the fiber noise contributed by arms A and B to be equal in magnitude, with coefficients derived from Table \ref{tab:coeffs}, but uncorrelated. Similarly, for scenarios 6 and 7 we assumed local lasers noise to be equal in magnitude but uncorrelated.

\renewcommand{\glossarymark}[1]{}
\printglossary
\printnoidxglossary[type=\acronymtype] 

%\bibliography{QKDphase}


\end{document}